\documentclass[acmsmall,screen,nonacm]{acmart}

\AtBeginDocument{  }

\setcopyright{none}
\renewcommand\footnotetextcopyrightpermission[1]{}
\usepackage{booktabs}                           \usepackage{subcaption}                         
\usepackage[utf8]{inputenc}
\usepackage[T1]{fontenc}
\usepackage[scaled=0.78]{beramono}
\usepackage{amsmath}
\usepackage{color}
\usepackage[dvipsnames]{xcolor,colortbl}
\usepackage{listings}
\usepackage{paralist}
\usepackage[font={small}]{caption}
\usepackage{wrapfig}
\usepackage{enumitem}
\usepackage{multicol}
\usepackage{multirow}
\usepackage{flushend}
\usepackage{bcprules}
\usepackage{textcomp}
\usepackage{tikz}
\usetikzlibrary{positioning,fit,calc,arrows.meta,arrows,decorations,shapes,tikzmark,backgrounds,overlay-beamer-styles}
\usetikzlibrary{decorations.pathreplacing,shapes.misc}
\usetikzlibrary{matrix}
\usepackage{pdfpages}
\usepackage{cleveref}
\usetikzlibrary{matrix}
\usetikzlibrary{tikzmark}
\usepackage{xspace}
\definecolor{light-gray}{gray}{0.85}
\usepackage{stackengine}
\usepackage{mdframed}
\usepackage{mathtools}
\usepackage{algorithm}
\usepackage[noend]{algpseudocode}
\usepackage{xurl}
\usepackage{bbold}
\usepackage{makecell}

\definecolor{ckeyword}{HTML}{000000}
\definecolor{ccomment}{HTML}{3F7F5F}
\definecolor{cstring}{HTML}{2A0099}

\lstdefinelanguage{Wasm}%
{alsoletter={.},%
  morekeywords={ module, func, param, result, export, local.get, local.set, i32,
  i64, f32, f64, i32.add, i32.sub, i32.mul, i32.div_s, i32.div_u, i32.rem_s,
  i32.rem_u, i32.and, i32.or, i32.xor, i32.shl, i32.shr_s, i32.shr_u, i32.gt_s, i64.add,
  i64.sub, i64.mul, i64.div_s, i64.div_u, i64.rem_s, i64.rem_u, f32.add,
  f32.sub, f32.mul, f32.div, f64.add, f64.sub, f64.mul, f64.div, if, then, else,
  end, block, loop, br_if, br, call, return, start },%
  sensitive=true,% case-sensitive keywords
  moredelim=*[il][\bfseries]{\#\#\ },
  morecomment=[l];,% single-line comments
  morecomment=[s]{/*}{*/},% multi-line comments
  morestring=[b]",% double-quoted strings
  showstringspaces=false%
}[keywords,comments,strings]%

\lstdefinelanguage{Scala}%
{morekeywords={
  abstract, sealed, lazy,
  case,catch,char,class,%
  def,do,else,extends,final,finally,for,%
  if,import,implicit,%
  match,module,%
  new,null,undefined,%
  array,
  override,%
  package,private,protected,public,%
  for,public,return,super,%
  this,throw,trait,try,type,%
  val,var,%
  with,while,%
  object,
  let,in,skip,assert,then,fst,snd,idx,sum,prod,exists,forall,%
  yield,%
  define, null?, car, cdr
  },%
  sensitive,%
  moredelim=*[il][\bfseries]{\#\#\ },
  morecomment=[l]//,%
  morecomment=[s]{/*}{*/},%
  morestring=[b]",%
  showstringspaces=false%
}[keywords,comments,strings]%

\lstdefinelanguage{Effect}%
{morekeywords={
    effect, yield, return
  },%
  sensitive,%
  moredelim=*[il][\bfseries]{\#\#\ },
  morecomment=[l]//,%
  morecomment=[s]{/*}{*/},%
  morestring=[b]",%
  showstringspaces=false%
}[keywords,comments,strings]%

\lstdefinelanguage{LLVMCPP}%
{morekeywords={
    define, i32, br, icmp, sub, call, mul, phi, ret, label, int, return,
    using, void, function, if, else, Cont, List
  },%
  sensitive,%
  moredelim=*[il][\bfseries]{\#\#\ },
  morecomment=[l]//,%
  morecomment=[s]{/*}{*/},%
  morestring=[b]",%
  showstringspaces=false%
}[keywords,comments,strings]%

\lstdefinelanguage{CPP}%
{morekeywords={
    using, void, function, if, else, Cont, List, return, int
  },%
  sensitive,%
  moredelim=*[il][\bfseries]{\#\#\ },
  morecomment=[l]//,%
  morecomment=[s]{/*}{*/},%
  morestring=[b]",%
  showstringspaces=false%
}[keywords,comments,strings]%

\lstdefinestyle{small}{
  language=Scala,%
  mathescape=true,%
  aboveskip=2pt,%\smallskipamount,
  belowskip=1pt,%\negsmallskipamount,
  lineskip=-1pt,
  basewidth={0.6em, 0.45em},%
  basicstyle=\fontsize{7}{9}\selectfont\ttfamily,
  keywordstyle=\keywordstyle,
  commentstyle=\commentstyle,
  stringstyle=\stringstyle,
  literate={-->}{{$\to$}}3
           {=>}{{$\Rightarrow ~$}}2
           {|-}{{$\ts$}}2
           {idx}{{$\#$}}1
           {sum}{{$\Sigma$}}1
           {array(}{{$\langle.\rangle$(}}3
           {σ}{{$\sigma$}}1
           {ρ}{{$\rho$}}1
           {→}{{$\to$}}1
           {λ}{{$\lambda$}}1
           {α}{{$\alpha$}}1
           {⊔}{{$\sqcup$}}1
           {⊓}{{$\sqcap$}}1
           {⊑}{{$\sqsubseteq$}}1
           {⊤}{{$\top$}}1
           {⊥}{{$\bot$}}1
           {×}{{$\times$}}1
           {τ}{{$\tau$}}1
           {ψ}{{$\psi$}}1
}

\lstdefinestyle{extrasmall}{
  language=Scala,%
  mathescape=true,%
  aboveskip=2pt,%\smallskipamount,
  belowskip=1pt,%\negsmallskipamount,
  lineskip=-1pt,
  basewidth={0.6em, 0.45em},%
  basicstyle=\fontsize{6}{8}\selectfont\ttfamily,
  keywordstyle=\keywordstyle,
  commentstyle=\commentstyle,
  stringstyle=\stringstyle,
  literate={-->}{{$\to$}}3
           {->}{{$\mapsto$}}3
           {=>}{{$\Rightarrow ~$}}2
           {|-}{{$\ts$}}2
           {idx}{{$\#$}}1
           {sum}{{$\Sigma$}}1
           {array(}{{$\langle.\rangle$(}}3
           {σ}{{$\sigma$}}1
           {ρ}{{$\rho$}}1
           {→}{{$\to$}}1
           {λ}{{$\lambda$}}1
           {α}{{$\alpha$}}1
           {⊔}{{$\sqcup$}}1
           {⊓}{{$\sqcap$}}1
           {⊑}{{$\sqsubseteq$}}1
           {⊤}{{$\top$}}1
           {⊥}{{$\bot$}}1
           {×}{{$\times$}}1
}

\definecolor{listingbg}{RGB}{240, 240, 240}

\newcommand{\commentstyle}[1]{\color{ccomment}\itshape{#1}}
\newcommand{\keywordstyle}[1]{\color{ckeyword}\bfseries{#1}}
\newcommand{\stringstyle}[1]{\color{cstring}\text{#1}}

\lstnewenvironment{listing}{\lstset{language=Scala}}{}
\lstnewenvironment{listingtiny}{\lstset{language=Scala,basicstyle=\scriptsize\ttfamily}}{}

\newcommand{\code}[1]{\lstinline[language=Scala,columns=fixed,basicstyle=\ttfamily]|#1|}

\newcommand{\den}[1]{\llbracket~#1~\rrbracket}

\newcommand{\silent}[1]{}

\newcommand{\tool}{\textsc{GenWasym}\xspace}
\newcommand{\GenSym}{\textsc{GenSym}\xspace}
\newcommand{\SymCC}{\textsc{SymCC}\xspace}
\newcommand{\QSYM}{\textsc{Qsym}\xspace}
\newcommand{\WASP}{\textsc{WASP}\xspace}
\newcommand{\Zthree}{\textsc{Z3}\xspace}
\newcommand{\Owi}{\textsc{Owi}\xspace}
\newcommand{\SeeWasm}{\textsc{SeeWasm}\xspace}
\newcommand{\Manticore}{\textsc{Manticore}\xspace}
\newcommand{\TO}{\textsc{TO}}

\newcommand{\Typ}[1]{\ensuremath{\mathsf{#1}}}

\newcommand{\lang}{$\mu$\textsf{Wasm}\xspace}

\newcommand\mydots{\ifmmode\ldots\else\makebox[1em][c]{.\hfil.\hfil.}\thinspace\fi}

\newcommand{\HL}[2][teal!12]{\ensuremath{\mathchoice%
  {\setlength{\fboxsep}{.5ex}\colorbox{#1}{$\displaystyle#2$}}%
  {\setlength{\fboxsep}{.5ex}\colorbox{#1}{$\textstyle#2$}}%
  {\setlength{\fboxsep}{.5ex}\colorbox{#1}{$\scriptstyle#2$}}%
  {\setlength{\fboxsep}{.5ex}\colorbox{#1}{$\scriptscriptstyle#2$}}}}%

\newcommand{\Ans}{\ensuremath{\mathsf{Ans}}}

\newcommand{\Cont}{\ensuremath{\mathsf{Cont}}}

\newcommand{\Cdeno}[1]{\ensuremath{ \mathbb{E}\den{#1} }}

\newcommand{\Concdeno}[1]{\ensuremath{ \mathbb{C}\den{#1} }}

\newcommand{\SConcdeno}[1]{\ensuremath{ \mathbb{S}\den{#1} }}
\newcommand{\MConcdeno}[1]{\ensuremath{ \mathbb{M}\den{#1} }}

\newcommand{\Pow}[1]{\ensuremath{\mathcal{P}(#1)}}

\newcommand{\ODeno}[1]{\ensuremath{ \mathbb{O}\den{#1} }}

\newcommand{\truncate}[2]{\ensuremath{\left\lfloor #1 \right\rfloor_{#2}}}
\newcommand{\concat}{\ensuremath{+\mkern-13mu+}\xspace}
\newcommand{\magic}{\hspace{-21em}}
\newcommand{\pconcat}[1]{\ensuremath{\prescript{}{#1}+\mkern-13mu+}\xspace}

\newcommand{\Quote}[1]{\ensuremath{\texttt{\textquotesingle}\{\, #1 \,\}}}
\newcommand{\Quo}[1]{\ensuremath{\texttt{\textquotesingle}#1}}
\newcommand{\QuoteT}[1]{\ensuremath{\Typ{Expr}[#1]}}

\newcommand{\Splice}[1]{\ensuremath{\texttt{\$} #1 }}
\newcommand{\Splicee}[1]{\ensuremath{\texttt{\$}\{\, #1 \,\}}}
\newcommand{\SV}[1]{\textcolor{RedOrange}{#1}}  % variable bound to static value
\newcommand{\DV}[1]{\textcolor{NavyBlue}{#1}} % variable bound to next-stage value
\newcommand{\qspace}{\phantom{\texttt{\textquotesingle}\{\, }}

\newcommand{\run}{\ensuremath{\mathsf{run}}}

\newcommand{\Stack}{\Typ{Stack}}
\newcommand{\Env}{\Typ{Env}}

\newcommand{\Symbolic}{\Typ{Symbolic}}

\newcommand{\PC}{\Typ{PC}}

\newcommand{\UNSAT}{\Typ{UNSAT}}

\algnewcommand{\IIf}[1]{\State\algorithmicif\ #1\ \algorithmicthen}
\algnewcommand{\EndIIf}{\unskip\ \algorithmicend\ \algorithmicif}

\newcommand{\ft}{\mathit{ft}}
\newcommand{\es}{\mathit{es}}
\newcommand{\Inst}[1]{\ensuremath{\mathsf{#1}}}

\newcommand{\lift}{\ensuremath{\mathsf{lift}}}
\newcommand{\take}{\ensuremath{\mathsf{take}}}
\newcommand{\pop}{\ensuremath{\mathsf{pop}}}
\newcommand{\tail}{\ensuremath{\mathsf{tail}}}
\newcommand{\emptymap}{\ensuremath{\varnothing}}

\newcommand{\memo}{\ensuremath{\mathit{memo}}}

\lstdefinelanguage{DOT}{morekeywords={val,new},  sensitive,  morecomment=[l]//,  morecomment=[s]{/*}{*/},  morestring=[b]",  morestring=[b]',  showstringspaces=false}[keywords,comments,strings]

\newlength{\trulemargin}
\newlength{\trulewidth}
\newlength{\srulewidth}
\newenvironment{trules}{$\vspace{0.5em}\ba{p{\trulemargin}@{~}p{\trulewidth}@{~}p{\trulemargin}}}{\ea$}
\newenvironment{srules}{$\vspace{0.5em}\ba{p{\trulemargin}@{~}p{\srulewidth}}}{\ea$}

\newcommand{\ba}{\begin{array}}
\newcommand{\ea}{\end{array}}

\newcommand{\ei}{\end{array}}
\newcommand{\bcases}{\left\{\begin{array}{ll}}
\newcommand{\ecases}{\end{array}\right.}

\newcommand{\ts}{\,\vdash\,}

\newcommand{\la}{\langle}
\newcommand{\ra}{\rangle}
\newcommand{\eg}{{\em e.g.}\xspace}
\newcommand{\ie}{{\em i.e.}\xspace}

\newcommand{\judgement}[2]{{\textsf{\textbf{#1}}} \hfill #2}

\tikzset{
  invisible/.style={opacity=0,text opacity=0},
  visible on/.style={alt=#1{}{invisible}},
  alt/.code args={<#1>#2#3}{\alt<#1>{\pgfkeysalso{#2}}{\pgfkeysalso{#3}}}
}

\begin{document}

\title[]{Compiling WebAssembly Concolic Execution with Staging, Continuations, and Snapshots (Extended Version)}
\titlenote{ 
  This paper is an extended version of the paper of the same title published at
  OOPSLA 2026. Its appendix includes complete definitions of the formal semantics and
  algorithms.
}

\author{Dinghong Zhong}
\orcid{0009-0005-6280-1692}
\affiliation{
  \institution{Tufts University}
  \city{Medford}
  \country{USA}
}
\email{dinghong.zhong@tufts.edu}

\author{Alexander Y. Bai}
\orcid{0009-0009-7458-7864}
\affiliation{
  \institution{New York University}
  \city{New York}
  \country{USA}
}
\email{alexander.bai@nyu.edu}

\author{Mikail Khan}
\orcid{0009-0002-0646-9064}
\affiliation{  
  \institution{Carnegie Mellon University}
  \city{Pittsburgh}
  \country{USA}
}
\email{mikailk@cmu.edu}

\author{Guannan Wei}
\correspondingauthor
\orcid{0000-0002-3150-2033}
\affiliation{  \institution{Tufts University}
  \city{Medford}
  \state{MA}
  \country{USA}
}
\email{guannan.wei@tufts.edu}

\begin{abstract}
  Concolic execution is a variant of symbolic execution that runs a program
  simultaneously with concrete and symbolic inputs.
  It records the symbolic constraints encountered along a concrete execution
  path, then solves those constraints to generate inputs that explore new paths.
  Existing concolic engines generally follow one of two implementation strategies:
  Interpreter-based systems are comparatively simple to build but incur
  substantial interpretation overhead, while instrumentation-based systems
  avoid this overhead but typically re-execute the program from the beginning
  for each new input.

  In this paper, we develop a new approach that achieves the best of both worlds.
  Starting from the concrete semantics of the target language, we first develop a
  definitional concolic interpreter and stage it to compile away
  interpretation overhead while retaining the simplicity of an
  interpretation-based implementation.
  By expressing the staged interpreter in continuation-passing style, we can
  capture execution snapshots at branch points and resume from them when
  exploring alternative paths, avoiding repeated execution from the program
  entry.
  Because snapshot-reuse can itself incur overhead, we further develop a
  heuristic that favors snapshot-reuse only when it is expected to be
  beneficial.
  We instantiate this approach for WebAssembly and implement it in a new
  concolic-execution compiler \tool. Across 184 benchmarks, \tool with staging
  alone achieves a $29.4\times$ average speedup over the interpreter-based
  \WASP; heuristic snapshot-reuse further increases the speedup to
  $44.9\times$.
\end{abstract}

%% 2012 ACM Computing Classification System (CSS) concepts
%% Generate at 'http://dl.acm.org/ccs/ccs.cfm'.

%% End of generated code

%% Keywords
%% comma separated list
  %% \keywords are mandatory in final camera-ready submission

\maketitle
\section{Introduction}
As a white-box testing technique, modern symbolic execution (SE)
\cite{Cadar:2013:SES:2408776.2408795} automates test generation and bug finding
by partitioning the input space of programs and utilizing the power of SMT
solvers~\cite{barrett2021satisfiability}.  By treating some inputs as symbolic
values \cite{Boyer:1975:SFS:800027.808445, King:1976:SEP:360248.360252, 1702443,
DBLP:conf/acm/Clarke76}, symbolic execution calculates the conditions of
program paths symbolically. These conditions represent equivalence classes of
inputs that can trigger the execution of specific paths.
General nondeterministic symbolic execution (\eg, KLEE
\cite{10.5555/1855741.1855756}, GenSym \cite{icse23}) forks the execution and
explores multiple paths simultaneously.
However, this approach requires frequent path feasibility
checks and quickly leads to path explosion.

Concolic execution \cite{10.1145/1065010.1065036, DBLP:conf/sigsoft/SenMA05,
DBLP:conf/ndss/GodefroidLM08} alleviates solver overhead and path explosion by
constraining itself to run only one path at a time, guided by a
concrete input.
Since path feasibility is evidenced by that concrete input, it
suffices to collect the corresponding path condition without querying solvers
when running the program.
In this way, concolic execution mixes \textbf{conc}rete and symb\textbf{olic} execution.
After executing one path, new inputs are generated by negating parts of the
collected path condition and solving satisfiability for the new concrete inputs.
These new inputs lead the engine to explore new paths in the program.
A driver orchestrates this process iteratively to systematically explore the program space.
In practice, concolic execution is often more effective for bug finding and is
the building block of widely adopted hybrid fuzzing techniques
\cite{DBLP:conf/icse/MajumdarS07, DBLP:conf/uss/Yun0XJK18,
DBLP:conf/ndss/StephensGSDWCSK16}, as concrete inputs can drive execution deep
into the program.

\subsubsection*{\textbf{Interpretation or Instrumentation?}}
Concolic execution can be viewed as a specific ``path scheduling strategy'', where
the path to be explored is biased by the concrete inputs provided in each iteration.
However, in each iteration with a new input, concolic execution needs
to rerun the program from the beginning (\ie, the root of the symbolic execution tree),
both concretely and symbolically. Although a prefix of the execution is shared
between the paths explored by the new and old inputs, the
corresponding path conditions for the new path must still be re-collected through
unnecessary re-execution. This redundant re-execution hurts performance
and throughput of concolic execution, especially when the negation point lies
deep in the execution tree.

Indeed, re-execution can be avoided by reusing the existing symbolic state
from the negation point, an optimization known as \emph{snapshot reuse}
(\eg, Driller \cite{DBLP:conf/ndss/StephensGSDWCSK16}).
It has been observed that Driller with snapshot reuse achieves more coverage in
76 out of 126 benchmarks from DARPA's Cyber Grand Challenge \cite{DBLP:conf/uss/Yun0XJK18}.
Realizing snapshot reuse requires copying the program state and saving it at the
corresponding branching point, so that in the next iteration, execution can
directly jump back to that point and continue the alternative branch.
Due to these requirements, implementing snapshot reuse is most straightforward
within interpretation-based concolic execution,
where the engine has full control
over the execution state and control flow at the meta level.
However, copying the whole program state can be expensive \cite{DBLP:conf/uss/Yun0XJK18}, and interpretation imposes
significant overhead \cite{DBLP:books/daglib/0072559}, such as traversing and
dispatching over the program's static AST/IR. These overheads make interpretation a less
ideal choice for performance-critical concolic execution, even with snapshot
reuse.

To achieve high performance, state-of-the-art concolic execution engines (\eg,
\textsc{QSYM} \cite{DBLP:conf/uss/Yun0XJK18} for binaries, \textsc{SymCC}
\cite{poeplau2020symcc} for LLVM IR, and \textsc{SymQEMU} \cite{EURECOM+6459} for
VEX IR) rely on source-to-source \emph{instrumentation} to completely avoid interpretation overhead.
Since concolic execution explores only one path at a time, instrumentation needs
only to insert symbolic instructions that collect path conditions alongside
the control flow of the original concrete execution.
However, these instrumentation-based engines abandon snapshot reuse at all.
It is no coincidence: with instrumentation that shallowly inserts additional
instructions, the program itself cannot save its current control at a
branching point
\footnote{To be precise, it would require the language to have mechanisms such as
\texttt{call/cc} to capture the current continuation, which is not the case here for
typical targets of concolic execution.} and later resume it in the next
iteration.
Therefore, it is challenging to implement snapshot reuse entirely within
instrumentation-based concolic execution.

\subsubsection*{\textbf{Best of Two Worlds}}
Although instrumentation enables fast execution, snapshot reuse remains a valuable
optimization, especially when branch points occur deep in the execution.
To maximize throughput, these two techniques should be combined.
Existing implementation strategies, however, force concolic-execution engine designers
to choose between them: snapshot reuse is relatively straightforward to implement
in an interpreter, but incurs substantial interpretation overhead;
instrumentation eliminates that overhead, but makes snapshot reuse difficult to realize.

In this paper, we resolve this tension by showing how to obtain instrumentation-like
performance while still supporting effective snapshot reuse.
Our key idea is to partially evaluate \cite{DBLP:books/daglib/0072559} a concolic
interpreter written in continuation-passing style (CPS) \cite{DBLP:journals/lisp/Reynolds93, appel1992compiling}
using staging \cite{DBLP:conf/pepm/TahaS97, taha1999multi, DBLP:conf/gpce/RompfO10}.

\begin{itemize}[leftmargin=1.5em]
\item By staging a concolic interpreter, we specialize it with respect to
  the input program and thereby generate efficient concolic-execution code that
  no longer suffers from interpretation overhead, achieving instrumentation-like performance.
  In effect, our approach realizes compilation via the first Futamura
  projection \cite{Futamura1971, Futamura1999} for a concolic interpreter.
  At the same time, constructing a concolic interpreter is arguably simpler than building
  a compiler, so our approach reduces the implementation effort required for high-performance engines.

\item By writing the staged concolic interpreter in CPS, we make control flow explicit
  and manipulable as first-class functions in the generated code, which enables snapshot reuse.
  At each branch in the source program, we generate code that stores the current
  state and continuation as a snapshot.
  When execution later resumes from that branch point, the engine invokes the saved
  continuation for the alternative branch, completely avoiding re-execution from
  the program entry to that point.
\end{itemize}

Applying staging or partial evaluation to program analysis is not new
\cite{damian1999partial, amtoft1999partial, Wei:2019:SAI:3366395.3360552,
DBLP:conf/cc/BoucherF96, 10.1145/3428232, icse23}, nor is the use of continuations
in symbolic execution engines \cite{icse23}.
Our work is inspired in part by GenSym's metaprogramming approach \cite{icse23},
which shows that, for nondeterministic symbolic execution of LLVM IR, continuations
can be used to implement search heuristics.
We adopt the idea of using continuations to manipulate control flow at the meta level.
To the best of our knowledge, however, this is the first work to combine staging and
continuations in a concolic execution engine to achieve both high execution performance
and snapshot reuse, unifying two optimizations that have previously been deployed only
separately in practice.

That said, the effectiveness of snapshot reuse still depends on the target program.
When symbolic states are large and execution paths are shallow, saving and restoring
snapshots can cost more than simply re-executing from the beginning, since the avoided
re-execution may be minimal. To address this trade-off, we propose a heuristic that
decides when to use snapshots based on execution depth and symbolic-state complexity
(\Cref{sec:heuristic}).
Our heuristic also avoids using snapshots for external functions, which
may have side effects depending on the concrete inputs.
With this heuristic, our approach smoothly combines the best of both worlds:
\emph{efficient execution through compilation and adaptive snapshot reuse via continuations
when it is likely to pay off}.
\subsubsection*{\textbf{Compiled Concolic Execution with Snapshots for WebAssembly}}
To demonstrate and validate our approach, we use WebAssembly (Wasm)
\cite{DBLP:conf/pldi/HaasRSTHGWZB17} as the target language.
As a general-purpose, low-level bytecode language, WebAssembly has emerged as an important
target for both compilation and program analysis.
WebAssembly is now supported by all major browsers and has also been adopted beyond
the browser setting.

Despite WebAssembly's growing importance,
existing concolic or symbolic execution tools for Wasm remain interpreter-based,
limiting their performance and scalability.
To the best of our knowledge, there is no instrumentation- or compilation-based
concolic execution engine for Wasm.
Therefore, our focus on Wasm is motivated both by its growing importance as a
deployment and analysis target and by the current lack of high-performance concolic
execution tools for it.

We present our development through a sequence of mechanical derivation steps, each obtained
from the previous one. We begin with a concrete continuation-passing semantics for
WebAssembly (\Cref{sec:background}), formulated as a definitional interpreter \cite{Reynolds:1972aa}
adapted from \cite{DBLP:conf/sfp/WeiBZZ25}.
On top of this foundation, we derive a concolic evaluator in CPS
(\Cref{sec:concolic}), refactor it into a staged concolic
interpreter in CPS (\Cref{sec:ce-compiler}), and finally show how snapshots can be reused
in the generated code (\Cref{sec:timetravel}).

Although this paper focuses on WebAssembly as the target language, our approach is not specific
to WebAssembly and readily extends to other languages or intermediate representations (\Cref{sec:discussion}).
Our mechanical, derivational methodology helps ensure that each step is correct by
construction and makes the overall approach easier to transfer to other languages.

\subsubsection*{\textbf{Implementation and Evaluation}}

We implement our architecture in \tool, a concolic-execution compiler for Wasm
written in Scala using LMS \cite{DBLP:conf/gpce/RompfO10}. Given a Wasm program,
\tool generates a C++ implementation following our snapshot-reusing concolic semantics.
The generated program links against a scheduler that invokes the SMT solver to
generate new inputs and applies our heuristic (\Cref{sec:heuristic}), which chooses between
re-execution and applying snapshots. To the best of our
knowledge, \tool is the first concolic-execution compiler for Wasm and the first
Wasm concolic engine to support snapshot reuse.

We evaluate \tool against \WASP \cite{DBLP:conf/ecoop/MarquesS0A22},
the prior state-of-the-art concolic execution engine for WebAssembly.
\WASP suffers from two performance limitations
due to its design as an interpreter-based engine.
First, it is built on the reference small-step reduction semantics
\cite{DBLP:conf/pldi/HaasRSTHGWZB17}, which incurs substantial interpretation
cost when resolving control flow, such as jumps, through administrative
instructions \cite{wasmref-isabelle}.
Second, it does not support snapshot reuse, leading to redundant re-execution.
Our work addresses both limitations.

We evaluate \tool on the B-Tree benchmark suite and the Collections-C benchmark suite
\cite{srdja_collections_c} used by \WASP \cite{DBLP:conf/ecoop/MarquesS0A22}.
Across 184 benchmarks,
\tool with staging alone achieves an average (geomean) speedup of $29.4\times$.
Further enabling heuristic snapshot reuse yields an average
speedup of $44.9\times$.
The results show that staging uniformly eliminates the majority overhead from
interpretation, yielding orders-of-magnitude speedups, while on top of that
snapshot reuse provides significant additional improvement for
programs with deep execution paths.

\subsubsection*{\textbf{Contributions and Organization}}
In this paper, we develop the first concolic-execution architecture to combine
high-performance execution with snapshot reuse, enabled by the key insight of
using staging with continuations in the concolic-execution engine.
We demonstrate our approach with Wasm, implement the snapshot-reusing
concolic-execution compiler \tool, and evaluate its performance through
extensive experiments.
The paper is organized as follows:
\begin{itemize}[leftmargin=5mm]
  \item \Cref{sec:background} briefly reviews concolic execution, Wasm,
  and Wasm's concrete CPS semantics.
  \item \Cref{sec:concolic} develops a definitional concolic interpreter
    based on the concrete CPS semantics.
    \Cref{sec:ce-compiler} derives the staged concolic
    interpreter by refactoring the single-stage concolic interpreter.
  \item \Cref{sec:timetravel} explains how to implement snapshot reuse (\ie,
    avoiding re-execution) in the generated concolic execution code by
    manipulating continuations.
    \Cref{sec:heuristic} proposes a heuristic to decide when to reuse snapshots
    to balance the benefit and cost of snapshot resumption.
\item \Cref{sec:impl} describes our implementation of \tool, which is implemented
    in Scala and generates C++ code for concolic execution of WebAssembly programs.
  \item \Cref{sec:eval} presents the performance evaluation of \tool against
    WASP \cite{DBLP:conf/ecoop/MarquesS0A22}, showing an average $29.4\times$ speedup.
    We also evaluate the effectiveness of our snapshot heuristic.

\item \Cref{sec:discussion} compares our approach with instrumentation and
    discusses its applicability to other languages; \Cref{sec:related} reviews
    related work; and \Cref{sec:conclusion} concludes the paper.
\end{itemize}

\section{Background} \label{sec:background}

In this section, we provide background on concolic execution, WebAssembly, and
its CPS semantics~\cite{DBLP:conf/sfp/WeiBZZ25} that we build upon.

\subsection{Concolic Execution}

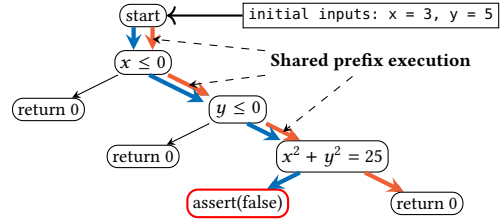
\begin{figure}[t]
\begin{minipage}[h]{0.49\textwidth}
\begin{lstlisting}[language=CPP]
int x = input();
int y = input();
if (x <= 0) return 0;
else if (y <= 0) return 0;
else if (x * x + y * y == 25)
  return assert(false);
else return 0;
\end{lstlisting}
\end{minipage}
\begin{minipage}[h]{0.49\textwidth}
  \scriptsize
  \begin{tikzpicture}[
    every node/.style={rectangle,draw,rounded corners,minimum width=2.5em,minimum height=1.2em,inner sep=2pt},
    level distance=0.62cm,
    level 1/.style={sibling distance=2.5cm},
    level 2/.style={sibling distance=2.5cm},
    level 3/.style={sibling distance=2.5cm},
    level 4/.style={sibling distance=2.5cm},
    edge from parent/.style={draw, -stealth}
  ]
  \node (root) {$\text{start}$}
  child { node (br-1) {$x \leq 0$}
  child { node (br-1-1){return 0} }
  child { node (br-1-2) {$y \leq 0$}
  child { node (br-1-2-1){return 0}}
  child { node (br-1-2-2) {$x^2 + y^2 = 25$}
  child { node[draw=red, thick, fill=white] (error) {assert(false)}}
  child { node (br-1-2-2-2) {return 0}}}} edge from parent[draw=none]};
  \draw[RedOrange,thick,ultra thick,-stealth,transform canvas={xshift=1.25mm}] (root) -- (br-1);
  \coordinate (m1) at ($(root)!0.5!(br-1)+(1.25mm,0)$);
  \draw[RedOrange,thick,ultra thick,-stealth] (br-1) -- (br-1-2) node[midway, coordinate] (m2) {};
  \draw[RedOrange,thick,ultra thick,-stealth] (br-1-2) -- (br-1-2-2) node[midway, coordinate] (m3) {};
  \draw[RedOrange,thick,ultra thick,-stealth] (br-1-2-2) -- (br-1-2-2-2);
  \node[rounded corners=0pt, right=of root, font=\scriptsize\ttfamily] (input) {initial inputs: x = 3, y = 5};
  \draw[->, thick] (input) -- (root);
  \draw[NavyBlue,thick,ultra thick,-stealth,transform canvas={xshift=-1.25mm}] (root) -- (br-1);
  \draw[NavyBlue,thick,ultra thick,-stealth,shorten >=-2mm,transform canvas={xshift=-2.5mm}] (br-1) -- (br-1-2);
  \draw[NavyBlue,thick,ultra thick,-stealth,shorten >=0.05mm,transform canvas={xshift=-2.5mm}] (br-1-2) -- (br-1-2-2);
  \draw[NavyBlue,thick,ultra thick,-stealth] (br-1-2-2) -- (error);
  \node[draw=none, below=1em of input,align=left](note) {\textbf{Shared prefix execution}};
  \draw[->, -stealth, dashed](note) -- (m1);
  \draw[->, -stealth, dashed](note) -- (m2);
  \draw[->, -stealth, dashed](note) -- (m3);
  \end{tikzpicture}
\end{minipage}
\vspace{-1em}
\caption{An example program (left) and its concolic execution tree (right).
The orange execution path is traced by initial inputs,
while the blue path represents another execution, sharing a prefix with the orange path.}
\label{fig:concolic:example}
\vspace{-1em}
\end{figure}

Concolic execution \cite{DBLP:series/natosec/BallD15, 10.1145/1065010.1065036,
DBLP:conf/sigsoft/SenMA05, DBLP:conf/ndss/GodefroidLM08} runs program $p$
with concrete input $i$, traces an execution path $\pi$, and collects its path
condition $\textit{cond}$. 
Solving negated path condition $\neg\textit{cond}$ with an SMT solver generates a
new input $i'$ that explores a another path. 
This process repeats until all paths are explored or a limit (\eg, time or
coverage) is reached.

Consider the example program and its execution tree shown in \Cref{fig:concolic:example}.
Suppose we start concolic execution with initial inputs $x=3$ and $y=5$.
The execution path in {\color{RedOrange}orange} collects the path condition $x >
0 \land y > 0 \land x^2 + y^2 \neq 25$ along the way.
To explore a new path, concolic execution negates the condition, \eg, $x > 0
\land y > 0 \land x^2 + y^2 = 25$ where the last condition is negated, and
queries the SMT solver for a new input satisfying the negated condition.
The solver returns a new input $x=4$ and $y=3$, which explores a new path shown
in {\color{NavyBlue} blue} in \Cref{fig:concolic:example}.
However, to reach the error node in the blue path, the concolic execution engine needs to
re-execute part of the program, which was already executed in the last
iteration.
If the underlying concolic execution engine can save the snapshot at
branch points, it can reuse those symbolic states to avoid re-execution.

The topic of this paper is to develop an approach to concolic execution that is
easy to implement, efficient to execute, and able to reuse symbolic states to
avoid re-execution.

\subsection{WebAssembly}

\begin{figure}\small
  \vspace{-0.5em}
  \begin{alignat*}{4}
    \ell & \in \Typ{Label} && \ \ = \mathbb{N}
      & \hspace{2em}x & \in \Typ{Identifier} && \ \ = \mathbb{N} \\
    t & \in \Typ{Value Type} && ::= \mathsf{i32} \mid \mathsf{i64} \mid \dots
      & \hspace{2em}\ft & \in \Typ{Function Type} && ::= t^* \rightarrow t^* \\
    e & \in \mathrlap{\Typ{Instruction} ::=
        \Inst{nop}
        \mid t.\Inst{const}~c
        \mid t.\{\Inst{add}, \Inst{sub}, \Inst{eq}, \dots\}
        \mid \Inst{local.get}~x
        \mid \Inst{local.set}~x} \\
      & \mathrlap{
      \phantom{\Typ{Instruction} ::=}\hspace{1em}
        \mid \Inst{block}~\ft~es
        \mid \Inst{loop}~\ft~es
        \mid \Inst{if}~\ft~es~es
        \mid \Inst{br}~\ell
        \mid \Inst{call}~x
        \mid \Inst{return}} \\
    \es & \in \Typ{Instructions} && = \Typ{List[Instruction]}
      & \hspace{2em}f & \in \Typ{Function} && ::= \Typ{func}~x~
          \{\Typ{type}:\ft,\ \Typ{locals}:t^*,\ \Typ{body}:\es\} \\
    m & \in \Typ{Module} && ::= \Typ{module}~f^*
  \end{alignat*}
  \vspace{-2em}
  \caption{The abstract syntax of \lang.}
  \label{fig:syntax}
  \vspace{-1.5em}
\end{figure}

We demonstrate our approach on WebAssembly, an emerging portable stack-based
virtual instruction set for safe and efficient execution on the web and beyond.
In the formalization, we limit our discussion to \lang, a subset of WebAssembly that
captures the core features relevant to concolic execution.
Readers may refer to the official specification \cite{wasmspec2} for
more details about WebAssembly.
\Cref{fig:syntax} shows the syntax of \lang, and \Cref{fig:concrete} shows its
concrete semantics.
\lang models the following instructions covering the essential part of local
variables, control flow, and function calls in WebAssembly:
\begin{itemize}[leftmargin=1.5em]
  \item We model a range of arithmetic instructions. For example,
    $t.\Inst{const} ~ c$ pushes a constant value $c$ of type $t$
    onto the stack, and $t.\Inst{add}$ pops two operands of type $t$ from the
    stack, then pushes the addition result back to the stack.
      Instruction $\Inst{local.get}~ x$ pushes the value of variable $x$ onto the
    stack, and $\Inst{local.set}~ x$ sets the value of variable $x$ using the
    top value on the stack.
  \item Unlike most other low-level IRs, WebAssembly supports structured
    control flow with \Inst{block}, \Inst{loop}, and \Inst{if} instructions, containing
    sequences of instructions as their bodies.
    Since block-like instructions can be nested, WebAssembly uses the $\Inst{br}~ l$
    instruction to transfer control-flow to the target block, where label $l$ is the
    de Bruijn index identifying the target block. Block-like instructions are annotated with a function type $t^m \to t^n$, indicating
    the shape of the stack before and after the execution of the block.
  \item A top-level program consists of a module of functions.
    Each function has an identifier, a type, a list of local variables, and a sequence of
    instructions as its body.
    The instruction $\Inst{call}~x$ transfers the control-flow to the callee,
    and $\Inst{return}$ transfers the control-flow back to the caller.
\end{itemize}
We omit typing of the language, which is standard and can be found in the
official WebAssembly specification \cite{wasmspec2}.
We also omit Wasm features such as tables, global variables, and
memory, which are not essential to explain our approach.
However, our implementation in \Cref{sec:impl} supports a larger subset of
WebAssembly and can handle more realistic programs.

\subsection{Continuation-Passing Semantics of \lang} \label{sec:cps-wasm}

\subsubsection*{Why Continuations?}
The official WebAssembly specification \cite{wasmspec2} defines a
small-step reduction semantics.
However, the semantics has to introduce administrative instructions to manage
control flow; these instructions represent evaluation contexts and are
materialized only at runtime.
This obscures the binding-time distinction between parts of
the program that are known at compile time and others at runtime, rendering it
difficult to construct a staged interpreter on top of it.

In this paper, we adopt a continuation-passing semantics
\cite{DBLP:conf/sfp/WeiBZZ25} for WebAssembly that is defined in a compositional
way using continuation functions to represent control flow.
It is well-known that continuations improve binding-time separation,
allowing more computation to be evaluated at compile time
\cite{DBLP:conf/lfp/LawallD94, DBLP:conf/pepm/HolstG91, DBLP:conf/fpca/ConselD91}.
We later exploit this insight and build the staged concolic interpreter on top
of the CPS semantics \cite{DBLP:conf/sfp/WeiBZZ25}.
Moreover, continuations naturally model the snapshots of symbolic states
needed for concolic execution, as we will discuss in \Cref{sec:timetravel}.

The correctness of the CPS semantics could be established by relating it to
the small-step semantics via defunctionalization \cite{DBLP:journals/scp/DanvyM09} and
refunctionalization \cite{DBLP:conf/ppdp/DanvyN01, DBLP:journals/programming/Gibbons22}.
A detailed investigation of this connection is out of the scope of this work.

\begin{figure}\small
\judgement{Semantic Domains}{}

\vspace{-1.5em}
\begin{minipage}[t]{0.49\textwidth}
  \begin{alignat*}{3}
    v      & \in \Typ{Value} && = \mathbb{Z} \\
    \sigma & \in \Typ{Stack} && = \Typ{List[Value]}
  \end{alignat*}
\end{minipage}
\begin{minipage}[t]{0.49\textwidth}
  \begin{alignat*}{3}
    \rho   & \in \Typ{Env}   && = \Typ{List[Value]} \\
    \kappa & \in \Typ{Cont}  && = \Typ{Stack} \times \Typ{Env} \to \Typ{Ans}
  \end{alignat*}
\end{minipage}\\[1ex]

\judgement{Concrete Continuation Semantics}{$\Cdeno{\cdot} : \Typ{List[Inst]} \rightarrow (\Typ{Stack} \times \Typ{Env} \times \Typ{Cont} \times \Typ{List[Cont]}) \rightarrow \Typ{Ans}$}
  \begin{alignat*}{3}
    & \Cdeno{\mathit{nil}} (\sigma, \rho, \kappa, \theta) && = \kappa(\sigma, \rho) \\
    & \Cdeno{\Inst{nop} :: \mathit{rest}} (\sigma, \rho, \kappa, \theta) && =
      \Cdeno{\mathit{rest}} (\sigma, \rho, \kappa, \theta) \\
    & \Cdeno{t.\Inst{const} ~ c :: \mathit{rest}} (\sigma, \rho, \kappa, \theta) && =
      \Cdeno{\mathit{rest}} (c :: \sigma, \rho, \kappa, \theta) \\
    & \Cdeno{t.\Inst{add} :: \mathit{rest}} (v_1 :: v_2 :: \sigma, \rho, \kappa, \theta) && =
      \Cdeno{\mathit{rest}} (v_1 + v_2 :: \sigma, \rho, \kappa, \theta) \\
    & \Cdeno {\Inst{local.get} ~ x :: \mathit{rest}} (\sigma, \rho, \kappa, \theta) && =
      \Cdeno{\mathit{rest}}(\rho(x) :: \sigma, \rho, \kappa, \theta) \\
    & \Cdeno{\Inst{local.set} ~ x :: \mathit{rest}}(v :: \sigma, \rho, \kappa, \theta) && =
      \Cdeno{\mathit{rest}}(\sigma, \rho[x \mapsto v], \kappa, \theta)\\
    & \Cdeno{\Inst{block} ~ (t^m \to t^n) ~ \es :: \mathit{rest}}(\sigma_{\mathit{arg}} \pconcat{m} \sigma, \rho, \kappa, \theta) && = \\
       & && \magic \Typ{let} ~ \kappa_1 \coloneqq \lambda (\sigma_1, \rho_1). \Cdeno{\mathit{rest}}(\truncate{\sigma_1}{n} \mathbin{+\!\!+} \sigma, \rho_1, \kappa, \theta) ~ \Typ{in} ~ \Cdeno{\es}(\sigma_{\mathit{arg}}, \rho, \kappa_1, \kappa_1 :: \theta) \\
    & \Cdeno{\Inst{loop} ~ (t^m \to t^n) ~ \es :: \mathit{rest}}(\sigma_{\mathit{arg}} \pconcat{m} \sigma, \rho, \kappa, \theta) && = \\
       & && \magic \Typ{let} ~ \kappa_1 \coloneqq \lambda (\sigma_1, \rho_1). \Cdeno{\mathit{rest}}(\truncate{\sigma_1}{n} \mathbin{+\!\!+} \sigma, \rho_1, \kappa, \theta) ~ \Typ{in} \\
       & && \magic \Typ{fix} ~ \kappa_2 \coloneqq \lambda (\sigma_2, \rho_2). \Cdeno{\es}(\truncate{\sigma_2}{m}, \rho_2, \kappa_1, \kappa_2 :: \theta) ~ \Typ{in} ~ \kappa_2(\sigma_{\mathit{arg}}, \rho) \\
    & \Cdeno{\Inst{if} ~ (t^m \to t^n) ~ \es_1 ~ \es_2 :: \mathit{rest}}(v :: \sigma_{\mathit{arg}} \pconcat{m} \sigma, \rho, \kappa, \theta) && = \\
       & && \magic \Typ{let} ~ es \coloneqq \Typ{if} ~ v \equiv 0 ~ \Typ{then} ~ \es_2 ~ \Typ{else} ~ \es_1 ~ \Typ{in} \\
       & && \magic \Typ{let} ~ \kappa_1 \coloneqq \lambda (\sigma_1, \rho_1). \Cdeno{\mathit{rest}}(\truncate{\sigma_1}{n} \mathbin{+\!\!+} \sigma, \rho_1, \kappa, \theta) ~ \Typ{in} ~ \Cdeno{es}(\sigma_{\mathit{arg}}, \rho, \kappa_1, \kappa_1 :: \theta) \\
    & \Cdeno{\Inst{br} ~ \ell :: \mathit{rest}}(\sigma, \rho, \kappa, \theta) && = \theta(\ell)(\sigma, \rho) \\
    & \Cdeno{\Inst{call} ~ x :: \mathit{rest}}(\sigma_{\mathit{arg}} \pconcat{m} \sigma, \rho, \kappa, \theta) && = \\
       & && \magic \Typ{let} ~ \{ \Typ{type}: t^m \to t^n, \Typ{locals}: \mathit{ts}, \Typ{body}: \es \} \coloneqq \Typ{lookupFunc}(x) ~ \Typ{in} \\
       & && \magic \Typ{let} ~ \rho_1 \coloneqq \mathsf{buildEnv}(\sigma_{\mathit{arg}}, \mathit{ts}) ~ \Typ{in} \\
       & && \magic \Typ{let} ~ \kappa_1 \coloneqq \lambda (\sigma_1, \rho_1). \Cdeno{\mathit{rest}}(\truncate{\sigma_1}{n} \concat \sigma, \rho, \kappa, \theta) ~ \Typ{in} ~ \Cdeno{\es}([], \rho_1, \kappa_1, [\kappa_1]) \\
    & \Cdeno{\Inst{return} :: \mathit{rest}}(\sigma, \rho, \kappa, \theta) && = \theta.\mathsf{last}(\sigma, \rho)
  \end{alignat*}
  \vspace{-2em}
  \caption{The concrete execution semantics of \lang \cite{DBLP:conf/sfp/WeiBZZ25}.}
  \label{fig:concrete}
  \vspace{-1.5em}
\end{figure}

\subsubsection*{CPS Semantics of \lang}
Figure \ref{fig:concrete} shows \lang's concrete CPS semantics \cite{DBLP:conf/sfp/WeiBZZ25}.
Values are represented as integers, and stack and environment are both lists of values.
We call the current stack and environment the program \emph{state}.
Continuations $\kappa$ are represented as functions from states to the final result \Ans.
The \Ans{} type is deliberately left abstract and can be instantiated with
different types, enabling the semantics to serve different purposes.

The semantics function $\Cdeno{\cdot}$ is recursively defined over the input list of instructions.
Given a list of instructions $es$,
$\Cdeno{\mathit{es}}$ returns a function taking a stack, environment, continuation,
and a trail, and produces \Ans.
If the list is empty, we simply invoke the continuation.
If the list is not empty, we pattern-match on the first instruction and
define its semantics accordingly, and recursively call $\Cdeno{\mathit{rest}}$
when necessary.
The trail is represented as a list of continuations and serves as a table,
modeling block jumps local to a function, which we will explain later with control flow.

In our semantics, we pass the updated state explicitly to recursive calls to
$\Cdeno{\cdot}$ or continuations rather than updating them in place.
The $\Inst{nop}$ instruction leaves the state unchanged, while $\Inst{const} ~
c$ pushes $c$ onto the stack.
$\Inst{local.get}~x$ loads the value at index $x$ in the environment onto the stack, while
$\Inst{local.set}~x$ pops the stack and writes the popped value at index $x$ in the environment.
For brevity, several rules pattern-match on the stack. For example, the
$t.\Inst{add}$ rule matches $v_1 :: v_2 :: \sigma$, where $v_1$ and $v_2$ are
the top two stack values and $\sigma$ is the remainder. We write $\sigma_1
\pconcat{m} \sigma_2$ for a stack split into a top segment $\sigma_1$ of length
$m$ and a remainder $\sigma_2$, and $\truncate{\sigma}{n}$ for the top $n$
values of $\sigma$.

\subsubsection*{Control Flow}
Now we go through several control-flow instructions to illustrate how
continuations and trail are used.
The control-flow instructions $\Inst{block}$, $\Inst{loop}$, and $\Inst{if}$ add
a new continuation to the control context $\theta$, which defines the behavior
of branching instruction $\Inst{br}$.
The $\Inst{block}$ instruction evaluates the body instructions $es$ with
an updated control context that adds a new continuation $\kappa_1$ to the trail $\theta$.
The continuation $\kappa_1$ captures the control flow after the block.

\begin{wrapfigure}[8]{r}[5.5em]{0.27\textwidth}
\small
\label{fig:br-example}
\begin{lstlisting}[language=Wasm, escapechar = !, xleftmargin=0.75em,xrightmargin=0pt,aboveskip=0pt,belowskip=0pt,framexrightmargin=0pt,lineskip=-2pt]
(loop
  !\tikz[remember picture] \node[anchor=base] (a) {};!...
  (block
    ...
    br 0 !\tikz[remember picture] \node[anchor=base] (b) {};!
    ...
   !\tikz[remember picture] \node[anchor=base] (d) {};!br 1)
  ... !\tikz[remember picture] \node[anchor=base] (c) {};!
 !\tikz[remember picture] \node[anchor=base] (e) {};!br 0)
\end{lstlisting}
\begin{tikzpicture}[remember picture, overlay]
  \draw[->, thick, gray] (b) to[out=-30,in=10,looseness=1] (c);
  \draw[->, thick, gray] (d) to[out=175,in=-150,looseness=0.65] (a);
  \draw[->, thick, gray] (e) to[out=175,in=-150,looseness=0.65] (a);
\end{tikzpicture}
\end{wrapfigure}
To enter this continuation via $\Inst{br} ~ l$ instruction,
we look up the $l$-th continuation in the trail $\theta$,
and the control flow will jump to the corresponding continuation. On the right, the first $\Inst{br} ~ 0$
jumps to the end of the innermost block, which is captured by the continuation at
the top of the trail \emph{within} the scope of the block.
The $\Inst{loop}$ instruction is similar to $\Inst{block}$, but the continuation $\kappa_2$ added to the
trail $\theta$ will re-evaluate the loop body $es$ when branched to via $\Inst{br} ~ l$.
This is exemplified by the $\Inst{br}~0$ at the end of the example.
The $\Inst{if}$ instruction first pops the value $v$ from the stack.
Then when $v$ is not zero, it recursively evaluates the instructions $es_1$ with
the continuation $\kappa_1$ pushes onto trail $\theta$, where $\kappa_1$
captures the control flow after the true branch. Otherwise it evaluates $es_2$
in the false-branch with the same control context.

\subsubsection*{Function Call and Return}
The $\Inst{call}~x$ instruction retrieves $x$'s definition using
$\mathsf{lookupFunc}$ and uses $\mathsf{buildEnv}$ to construct $\rho_1$ from
the argument list $\sigma_{\mathit{arg}}$, allocating extra space for local
variables.
Finally, the $\Inst{call}~x$ instruction evaluates the function body $es$ under
an empty stack, $\rho_1$ and a control context initialized with
continuation $\kappa_1$.
This control context corresponds to the target function's scope and is a
singleton list containing $\kappa_1$, which captures the control flow after the
function call.
In the function body, the $\Inst{return}$ instruction invokes the last trail continuation
$\kappa_1$, skipping all enclosing blocks.
Readers can refer to \cite{DBLP:conf/sfp/WeiBZZ25} for more details on Wasm's
CPS semantics.

\section{Concolic Execution with Continuations} \label{sec:concolic}

The concrete CPS semantics of \lang (\Cref{sec:cps-wasm}) forms the basis for
developing our compositional concolic execution semantics, which we present in
this section. 
When presenting formalization, we \HL{\text{highlight}} the symbolic extensions
relative to the concrete semantics.

\subsection{Concrete-Input-Controlled Symbolic Execution}

\begin{figure}\small
\judgement{Semantic Domains}{}
\begin{equation}
  \nonumber
  \begin{alignedat}{3}
    op^\#_2  & \in \{ +^\#, -^\#, \ldots \} && \hspace{2em} op^\#1 \in \{ -^\#, \dots \} \\
    x         & \in \HL{\Typ{SymVar}}        && := \Typ{Identifier} \\
    s         & \in \HL{\Typ{Symbolic}} && := v \mid \Typ{Sym}(x) \mid s_1 ~ op^\#_2 ~  s_2 \mid op^\#_1 ~ s_1\\
    \pi^\#    & \in \HL{\Typ{PC^\#}}       && = \Typ{List[Symbolic]} \\
  \end{alignedat}
  \qquad
  \begin{alignedat}{3}
    \sigma^\# & \in \Typ{Stack}^\# && = \Typ{List[Value \times \HL{\Typ{Symbolic}}]} \\
    \rho^\#   & \in \Typ{Env}^\# && = \Typ{List[Value \times \HL{\Typ{Symbolic}}]} \\
    \kappa^\# & \in \Typ{Cont^\#} && = (\Typ{Stack}^\# \times \Typ{Env}^\# \times \HL{\Typ{PC^\#}}) \to \Ans \\
    \theta^\# & \in \Typ{Trail^\#} && = \Typ{List[Cont^\#]}
  \end{alignedat}
\end{equation}
\renewcommand{\magic}{\hspace{-26.12em}}
\judgement{Concolic Continuation Semantics}{}
  \begin{alignat*}{3}
    & \Concdeno{\cdot} : \mathrlap{\Typ{List[Inst]} \rightarrow (\Typ{Stack^\#} \times \Typ{Env^\#}\times \HL{\Typ{PC^\#}} \times \Typ{Trail^\#} \times \Typ{Cont^\#}) \rightarrow \Typ{Ans}} \\
    & \Concdeno{t.\Inst{const} ~ c :: \mathit{rest}} (\sigma^\#, \rho^\#, \HL{\pi^\#}, \kappa^\#, \theta^\#) && \hspace{-4em} =
      \Concdeno{\mathit{rest}} (\la c, \HL{c}\ra :: \sigma^\#, \rho^\#, \HL{\pi^\#}, \kappa^\#, \theta^\#) \\
    & \Concdeno{t.\Inst{add} :: \mathit{rest}} (\la v_1, \HL{s_1}\ra :: \la v_2, \HL{s_2}\ra :: \sigma^\#, \rho^\#, \HL{\pi^\#}, \kappa^\#, \theta^\#) && \hspace{-4em} =
      \Concdeno{\mathit{rest}} (\la v_1 + v_2, \HL{s_1 ~ +^\# ~ s_2}\ra :: \sigma^\#, \rho^\#, \HL{\pi^\#}, \kappa^\#, \theta^\#) \\
    & \Concdeno {\Inst{local.get} ~ x :: \mathit{rest}} (\sigma^\#, \rho^\#, \HL{\pi^\#}, \kappa^\#, \theta^\#) && \hspace{-4em} =
      \Concdeno{\mathit{rest}}(\rho^\#(x) :: \sigma^\#, \rho^\#, \HL{\pi^\#}, \kappa^\#, \theta^\#) \\
    & \Concdeno{\Inst{local.set} ~ x :: \mathit{rest}}(\la v, \HL{s}\ra :: \sigma^\#, \rho^\#, \HL{\pi^\#}, \kappa^\#, \theta^\#) && \hspace{-4em} =
      \Concdeno{\mathit{rest}}(\sigma^\#, \rho^\#[x \mapsto \la v, \HL{s}\ra], \HL{\pi^\#}, \kappa^\#, \theta^\#)\\
    & \Concdeno{\Inst{if} ~ (t^m \to t^n) ~ \es_1 ~ \es_2 :: \mathit{rest}}(\la v, \HL{s}\ra :: \sigma^\#_{\mathit{arg}} \pconcat{m} \sigma^\#, \rho^\#, \HL{\pi^\#}, \kappa^\#, \theta^\#) && = \\
       & && \magic \Typ{let} ~ es \coloneqq \Typ{if} ~ v \equiv 0 ~ \Typ{then} ~ \es_2 ~ \Typ{else} ~ \es_1 ~ \Typ{in} \\
       & && \magic \Typ{let} ~ \HL{\pi^\#_1} \coloneqq \Typ{if} ~ v \equiv 0 ~ \Typ{then} ~ \neg s :: \pi^\# ~ \Typ{else} ~ s :: \pi^\# ~ \Typ{in} \\
       & && \magic \Typ{let} ~ \kappa^\#_1 \coloneqq \lambda (\sigma^\#_1, \rho^\#_1, \HL{\pi^\#_1}). \Concdeno{\mathit{rest}}(\truncate{\sigma^\#_1}{n} \mathbin{+\!\!+} \sigma^\#, \rho^\#_1, \HL{\pi^\#_1}, \kappa^\#, \theta^\#) ~ \Typ{in} \\
       & && \magic \Concdeno{es}(\sigma^\#_{\mathit{arg}}, \rho^\#, \HL{\pi^\#_1}, \kappa^\#_1, \kappa^\#_1 :: \theta^\#) \\
    & \Concdeno{\Inst{br} ~ \ell :: \mathit{rest}}(\sigma^\#, \rho^\#, \HL{\pi^\#}, \kappa^\#, \theta^\#) && \hspace{-4em} = \theta^\#(\ell)(\sigma^\#, \rho^\#,  \HL{\pi^\#})
  \end{alignat*}
  \vspace{-2em}
  \caption{The concolic semantics of \lang (excerpted). The complete semantics
    is in \Cref{sec:appendix:full-definitions}.}
  \label{fig:concolic:semantics}
  \vspace{-1.5em}
\end{figure}

We first extend the semantic domains with symbolic counterparts and then define
the concolic semantics as a definitional interpreter $\Concdeno{\cdot}$.
The new symbolic components collect path conditions during execution without
affecting the program's concrete behavior, so the concolic semantics subsumes
the concrete one. At the same time, it remains compositional and uses
continuations to make control flow explicit.

\subsubsection*{Symbolic Semantic Domains}
\Cref{fig:concolic:semantics} highlights the extended semantic domains.
$\Symbolic$ values $s$ are a superset of concrete values $v$, additionally
including symbols $\mathsf{Sym}(x)$ identified by $x$, and unary ($op^\#_1
~s_1$) and binary ($s_1 ~op^\#_2~ s_2$) symbolic expressions.

We use a single stack and environment to store both concrete and symbolic
values. Thus, the symbolic stack $\Stack^\#$ and environment $\Env^\#$ are lists
of pairs $\la v, s\ra$, where $v$ is a concrete value and $s$ is its corresponding
symbolic expression under the current path condition. Path conditions $\PC$ are
represented as lists of symbolic values, with newly encountered conditions
pushed to the front. The conjunction of all expressions in $\PC$ is the
constraints on symbolic inputs to trigger the current execution path.
Finally, continuation functions $\Cont^\#$ take a concolic stack, environment,
and path conditions as input. The return type $\Typ{Ans}$ is left
polymorphic to allow different instantiations.

\subsubsection*{Definitional Concolic Interpreter}
\Cref{fig:concolic:semantics} shows an excerpt of the concolic semantics as a definitional
interpreter. Compared to the concrete semantics (\Cref{fig:concrete}), the
concolic semantics $\Concdeno{\cdot}$ is extended to take path conditions
$\pi^\#$ as additional inputs, and other components are lifted to their concolic
counterparts.
Most of the changes to the interpreter are structural: the control flow is
unchanged and we simply pass around the symbolic components, therefore the cases
for $\mathit{nil}$, $\textsf{block}$, $\textsf{loop}$, $\textsf{call}$, and
$\textsf{return}$ are omitted and can be found in
\Cref{sec:appendix:full-definitions}.

We explain concolic execution through representative cases.
The symbolic interpretation of numeric operations mirrors their concrete construction.
For example, $t.\Inst{const}~c$ creates a concrete–symbolic pair $\la c, c\ra$ and pushes
it onto the concolic stack, while $t.\Inst{add}$ builds the corresponding
symbolic addition expression.
We collect path conditions when encountering conditional instructions.
In the $\Concdeno{\mathsf{if} \dots}$ case, we first pop the concolic condition
$\la v, s\ra$ from the stack.
The concrete value $v$ determines the branch to take, as in the concrete
semantics, while the symbolic value $s$ updates the path condition: depending on
the branch, $s$ or $\neg s$ is prepended to $\pi^\#$, yielding a new path
condition $\pi^\#_1$ that is passed to the evaluation of the chosen branch.

\subsubsection*{Initial Continuation}

To invoke the concolic definitional interpreter, we provide the initial halting
continuation $\kappa^\#_{halt}$. The interpreter's return type $\Typ{Ans}$ is
polymorphic over different instantiations of this continuation. For concolic
execution, we instantiate it to return the collected path conditions upon
completion $ \kappa^\#_{halt}(\_, \_, \pi^\#) = \pi^\# $.
The concolic execution scheduler (described below) uses the returned path
conditions to generate new inputs and explore alternative paths.

\subsection{Concolic Execution Scheduler}
Given an input Wasm program, the concolic scheduler (\Cref{fig:concolic:driver})
orchestrates the exploration of paths by invoking $\Concdeno{\cdot}$ with
varying inputs that lead to distinct paths.
The scheduler takes as input a module, the entry function name
$f$, and a list of initial arguments, which may include both symbolic (\ie,
$\Inst{Sym}(x)$) and concrete values.
It maintains a worklist $\mathit{WL} \in \Pow{\Typ{PC^\#}}$ of path conditions
to explore, where $\Pow{\cdot}$ is the powerset operator.
Each path condition represents constraints on symbolic inputs
leading to an unexplored path if satisfied.
Initially, the worklist contains only the empty path condition, since no paths
have yet been explored and any input is valid.

In each iteration, the scheduler pops a path condition $\mathit{pc}$ from the
worklist and invokes an SMT solver to find a satisfying assignment. We model the
solver as a function
$\mathit{solve} : \Typ{PC} \to \Typ{Model} \cup {\Typ{UNSAT}}$,
where a model is a partial mapping from symbolic variables to concrete values:
$$ m  \in \Typ{Model} = \Inst{SymVar} \rightharpoonup \Typ{Value}. $$
If $\mathit{pc}$ is unsatisfiable, the scheduler skips the path as there is
no possible input to reach this path.
Otherwise, to construct the initial concolic input stack for executing
$\Concdeno{\Inst{call}~f}$, we use an auxiliary function $\mathit{instantiate} : (\Typ{List[Symbolic]},
\Typ{Model}) \to \Typ{List[Value \times Symbolic]}$ that evaluates each symbolic
input under the model and pairs the resulting concrete value with the original
symbolic input.
The scheduler then invokes $\Concdeno{\Inst{call}~f}$ with the constructed
concolic stack, empty environment, initial continuation trail
$[\kappa^\#_{halt}]$, and halting continuation $\kappa^\#_{halt}$ to explore a
new path.

Upon completion, the interpreter returns the collected path condition $\pi_1$.
The scheduler then generates new path conditions by negating each constraint in
$\pi_1$, as detailed in \Cref{sec:appendix:full-definitions}, adds the resulting
conditions to the worklist, and proceeds to the next iteration.
When the worklist becomes empty, all feasible execution paths have been covered.

\begin{algorithm}[t]
  \small
  \caption{Concolic Scheduler}
  \label{fig:concolic:driver}
  \begin{algorithmic}[1]
    \Function{\textsc{Scheduler}}{$\mathit{mod} \in \Typ{Module}$, $f \in \Typ{Identifier}$, $\mathit{input} \in \Typ{List[Symbolic]}$}
      \State $WL \gets \{\, [] \, \}$ \Comment{$WL \in \Pow{\Typ{PC^\#}}$}
      \State $Seen \gets \varnothing$ \Comment{$Seen \in \Pow{\Typ{PC^\#}}$}
      \While{$WL \neq \varnothing$}
        \State pop $\pi$ from $WL$
        \IIf{$\pi \in Seen$} \textbf{continue}
        \State $m \gets \mathit{solve}(\pi)$ \Comment{invoke SMT solver}
        \IIf{$m = \UNSAT$} \textbf{continue} \Comment{unreachable path condition}
        \State $\pi_1 \gets \Concdeno{\Inst{call} ~ f}(\mathit{instantiate}(\mathit{input}, m), [], [\kappa^\#_{halt}], \kappa^\#_{halt})$ \Comment{run concolic semantics}
        \State $Seen \gets Seen \cup \{\pi_1\}$
        \ForAll{$\pi' \in \textsc{Negate}(\pi_1)$}
          \State $WL \gets WL \cup \{\pi'\}$
        \EndFor
      \EndWhile
    \EndFunction

    \Function{\textsc{Negate}}{$\pi \in \Typ{PC^\#}$} $\rightarrow$ $\Typ{List[PC^\#]}$ \dots \Comment{complete definition in the appendix}
    \EndFunction
  \end{algorithmic}
\end{algorithm}

\section{Compiling Concolic Execution with Continuations} \label{sec:ce-compiler}

The concolic semantics developed in \Cref{sec:concolic} can already be
implemented as a definitional interpreter. Although conceptually
straightforward, this interpreter incurs significant overhead from repeatedly
dispatching on instructions, which can greatly impact performance
\cite{DBLP:conf/apl/Wiedmann83,DBLP:conf/usenix/LionCSY22,DBLP:conf/pldi/WurthingerWHWSS17}.
In this section, we build a concolic-execution compiler to eliminate this
overhead by realizing the first Futamura projection
\cite{Futamura1999,Futamura1971} through multi-stage programming (MSP)
\cite{taha1999multi}. We refactor the concolic semantics
into a compiler that generates efficient concolic execution code without
interpretation overhead. \Cref{sec:compilation:msp} briefly introduces MSP and the
notations used later, and \Cref{sec:compilation:staging} presents the staged
concolic semantics for \lang.

\subsection{A Brief Overview of Multi-Stage Programming} \label{sec:compilation:msp}

Multi-stage programming is a generative programming paradigm that enables
programs to construct and manipulate other programs. By dividing computation
into multiple stages, programmers annotate code to indicate which parts depend
on static input known at earlier stages and which depend on dynamic input
available only later. The resulting residual program contains only computations
that depend on dynamic input, while computations over static input have been performed
in earlier stages. Writing an interpreter in an MSP language allows us to
specialize it to the static program representation, effectively achieving
programmer-guided partial evaluation \cite{DBLP:books/daglib/0072559}.

\begin{figure}\small
  \vspace{-1em}
  \[
  \begin{array}{@{}r@{\;}l@{\;}l@{\quad}l@{}}
    \Typ{fix}
      & \mathit{power}
      & : (\Typ{Int}, \Typ{Int}) \to \Typ{Int}
      & \coloneqq \lambda (x, n). \Typ{if} ~ n \equiv 0 ~ \Typ{then} ~ 1 ~ \Typ{else} ~ x * \mathit{power} ~ (x, n - 1) \\[2pt]
    \Typ{fix}
      & \mathit{genpower}
      & : (\Typ{\QuoteT{Int}}, \Typ{Int}) \to \Typ{\QuoteT{Int}}
      & \coloneqq \lambda (x, n). \Typ{if} ~ n \equiv 0 ~ \Typ{then} ~ \Quo{1} \\[-1pt]
    &&& \phantom{\coloneqq}\; \Typ{else} ~ \Quote{ \Splice{x} * \Splicee{\mathit{genpower} ~ (x, n - 1)} } \\[2pt]
    \Typ{let} & \mathit{power2}
      & : \Typ{Int} \to \Typ{Int}
      & \coloneqq \run ~ \Quote{\lambda (x). \Splicee{ \mathit{genpower}(\Quo{x}, 2) }}
  \end{array}
  \]
  \vspace{-1.5em}
  \caption{The example of unstaged power function, its staged version, and specialized
  power wrt $n = 2$.}
  \label{fig:staging:example}
  \vspace{-1em}
\end{figure}

Later, in \Cref{sec:compilation:staging}, we specify the staged concolic semantics
using the quasi-quotation and splicing syntax from Scala 3
\cite{DBLP:conf/gpce/StuckiBO21}.
A quotation $\Quote{e}$ (or simply $\Quo{e}$) denotes a next-stage expression
$e$, while splicing $\Quote{\dots \Splicee{e} \dots}$ inserts the value of a
current-stage expression $e$ into the surrounding quotation. At the type level,
$\Typ{\QuoteT{T}}$ represents a next-stage type $T$.

\Cref{fig:staging:example} illustrates MSP with the power function example. The
single-stage function $\mathit{power}$ computes $x^n$ recursively. When the
exponent $n$ is known in advance, we can write a multi-stage version
$\mathit{genpower}$, which generates a specialized $\mathit{power}$ function
taking only $x$ as input. In $\mathit{genpower}$, the variable $x$ is marked as
a next-stage value by type $\QuoteT{\Typ{Int}}$, and the result is a
next-stage expression that computes $x^n$. When $n = 0$, $\mathit{genpower}$
produces the staged constant $\Quo{1}$; when $n > 0$, it constructs a staged
expression multiplying $x$ by the result of a recursive call with $n - 1$.

By providing a concrete value for the static input $n$, we can specialize
$\mathit{genpower}$ and produce a residual program that depends only on $x$.
For instance, we specialize $\mathit{genpower}$ to obtain $\mathit{power2}$ by
applying $\run$, a common construct for evaluating staged expressions in MSP
languages \cite{DBLP:conf/gpce/StuckiBO21,10.1007/978-3-319-07151-0_6,DBLP:conf/pepm/TahaS97}.
Compared to the generic $\mathit{power}$
function, $\mathit{power2}$ for $n = 2$ unrolls all recursive calls and eliminates the overhead
of condition checks.

\subsection{Compilation of Concolic Semantics} \label{sec:compilation:staging}
We now develop the staged concolic semantics $\SConcdeno{\cdot}$ for \lang.
Specializing $\SConcdeno{\cdot}$ to a source program generates efficient
concolic execution code and realizes the first Futamura projection
\cite{Futamura1999,Futamura1971}.
Since the unstaged semantics $\Concdeno{\cdot}$ (\Cref{fig:concolic:semantics})
is compositional \cite{DBLP:conf/dagstuhl/Jones96}, we derive
$\SConcdeno{\cdot}$ in \Cref{fig:staging:semantics} by adding stage annotations
to $\Concdeno{\cdot}$ without altering the concolic evaluation logic.
During specialization, the staged semantics mostly unrolls recursive calls over subterms.

In the presentation of staged semantics, variables bound to current stage values are \SV{orange}, while
those bound to staged expressions are \DV{blue}.
We do not add any color for variables at next stage.

\renewcommand{\magic}{\hspace{-21.5em}}

\begin{figure}\small
\judgement{Staged Module Semantics}{}
  \begin{alignat*}{3}
    & \MConcdeno{\cdot} : \mathrlap{\Inst{List[Function]} \rightarrow \QuoteT{\Typ{Memo}}} \\
    & \MConcdeno{\mathit{nil}} && =  \emptymap \\
    & \MConcdeno{\{\Typ{type}: t^m \to t^n,  \Typ{id}: \SV{x},  \Typ{locals}: \SV{\mathit{ts}}, \Typ{body}: \SV{\es} \} :: \SV{\mathit{rest}}} && =\\
    & && \magic \Quote{\Typ{let} ~ f \coloneqq \lambda (\rho^\# , \pi^\# , \kappa^\#, m_1 ). \Splicee{\SConcdeno{\SV{\es}}(\Quo{m_1})(\lift([]), \Quo{\rho^\#}, \Quo{\pi^\#}, \Quo{\kappa^\#}, [ \Quo{\kappa^\#} ])} ~ \Typ{in} \\
    & && \magic \qspace \Quote{\MConcdeno{\SV{\mathit{rest}}}}[\Splice{\lift(\SV{x})} \mapsto f] }
  \end{alignat*}
  \vspace{-2em}
  \caption{The staged module semantics of \lang}
  \label{fig:staging:module}
  \vspace{-1em}
\end{figure}

\renewcommand{\magic}{\hspace{-22em}}

\begin{figure}\small

\judgement{Staged Concolic Continuation Semantics}{}
  \begin{alignat*}{3}
    & \SConcdeno{\cdot} : \mathrlap{\Typ{List[Inst]} \rightarrow \HL{\QuoteT{\Typ{Memo}}}} \\
    & \hspace{3.5em} \mathrlap{\rightarrow (\HL{\QuoteT{\Typ{Stack^\#}}} \times \HL{\QuoteT{\Typ{Env^\#}}}  \times \HL{\QuoteT{\Typ{PC^\#}}} \times \HL{\QuoteT{\Typ{Cont^\#}}} \times \HL{\Typ{Trail^\#}}) \rightarrow \HL{\QuoteT{\Typ{Ans}}}} \\
    & \SConcdeno{t.\Inst{const} ~ \SV{c} :: \SV{\mathit{rest}}}(\DV{\memo})(\DV{\sigma^\#}, \DV{\rho^\#}, \DV{\pi^\#}, \DV{\kappa^\#}, \SV{\theta^\#}) && =\\
       & && \magic
      \SConcdeno{\SV{\mathit{rest}}}(\DV{\memo})(\Quote{\la\Splice{\lift(\SV{c})}, \Splice{\lift(\SV{c})}\ra :: \Splice{\DV{\sigma^\#}}}, \DV{\rho^\#}, \DV{\pi^\#}, \DV{\kappa^\#}, \SV{\theta^\#}) \\
    & \SConcdeno{\Inst{block} ~ (t^m \to t^n) ~ \SV{\es} :: \SV{\mathit{rest}}}(\DV{\memo})(\DV{\sigma^\#}, \DV{\rho^\#}, \DV{\pi^\#}, \DV{\kappa^\#}, \SV{\theta^\#}) && = \\
      & && \magic \Quote{\Typ{let} ~ \kappa^\#_1 \coloneqq \lambda (\sigma^\#_1, \rho^\#_1, \pi^\#_1). \Splicee{\SConcdeno{\SV{\mathit{rest}}}(\Quote{\truncate{\sigma^\#_1}{n} \mathbin{+\!\!+} \Splice{\DV{\sigma^\#}}.\pop_m}, \Quo{\rho^\#_1}, \Quo{\pi^\#_1}, \DV{\kappa^\#}, \SV{\theta^\#})} ~ \Typ{in} \\
      & && \magic \qspace\Splicee{\SConcdeno{\SV{\es}}(\Quote{\Splice{\DV{\sigma^\#}}.\take_m}, \DV{\rho^\#}, \DV{\pi^\#}, \Quo{\kappa^\#_1}, \Quo{\kappa^\#_1} :: \SV{\theta^\#})} } \\
    & \SConcdeno{\Inst{loop} ~ (t^m \to t^n) ~ \SV{\es} :: \SV{\mathit{rest}}}(\DV{\memo})(\DV{\sigma^\#}, \DV{\rho^\#}, \DV{\pi^\#}, \DV{\kappa^\#}, \SV{\theta^\#}) && = \\
      & && \magic \Quote{\Typ{let} ~ \kappa^\#_1 \coloneqq \lambda (\sigma^\#_1, \rho^\#_1, \pi^\#_1). \Splicee{\SConcdeno{\SV{\mathit{rest}}}(\DV{\memo})(\Quote{\truncate{\sigma^\#_1}{n} \mathbin{+\!\!+} \Splice{\DV{\sigma^\#}}.\pop_m}, \Quo{\rho^\#_1}, \Quo{\pi^\#_1}, \DV{\kappa^\#}, \SV{\theta^\#})} ~ \Typ{in} \\
       & && \magic \qspace \Typ{fix} ~ \kappa^\#_2 \coloneqq \lambda (\sigma^\#_2, \rho^\#_2, \pi^\#_2). \Splicee{\SConcdeno{\SV{\es}}(\DV{\memo})(\Quote{\truncate{\sigma^\#_2}{m}}, \Quo{\rho^\#_2}, \Quo{\pi^\#_2}, \Quo{\kappa^\#_1}, \Quo{\kappa^\#_2} :: \SV{\theta^\#})} ~ \Typ{in} \\
       & && \magic \qspace \kappa^\#_2(\Splice{\DV{\sigma^\#}}.\take_m , \Splice{\DV{\rho^\#}}, \Splice{\DV{\pi^\#}}) } \\
    & \SConcdeno{\Inst{if} ~ (t^m \to t^n) ~ \SV{\es_1} ~ \SV{\es_2} :: \SV{\mathit{rest}}}(\DV{\memo})(\DV{\sigma^\#}, \DV{\rho^\#}, \DV{\pi^\#}, \DV{\kappa^\#}, \SV{\theta^\#}) && = \\
       & && \magic \Quote{\Typ{let} ~ \la v, s\ra \coloneqq \Splice{\DV{\sigma^\#}.\mathsf{head}} ~ \Typ{in} \\
       & && \magic \qspace\Typ{let} ~ \sigma^\#_{\mathit{arg}} \coloneqq \Splice{\DV{\sigma^\#}}.\mathsf{tail}.\take_m ~ \Typ{in} \\
       & && \magic \qspace\Typ{let} ~ \kappa^\#_1 \coloneqq \lambda (\sigma^\#_1, \rho^\#_1, \pi^\#_1). \Splicee{\SConcdeno{\SV{\mathit{rest}}}(\DV{\memo})(\Quote{\truncate{\sigma^\#_1}{n} \mathbin{+\!\!+} \Splice{\DV{\sigma^\#}}.\tail.\pop_m}, \Quo{\rho^\#_1}, \Quo{\pi^\#_1}, \DV{\kappa^\#}, \SV{\theta^\#})} ~ \Typ{in} \\
       & && \magic \qspace \Typ{let} ~ \pi^\#_1 \coloneqq \Typ{if} ~ v \equiv 0 ~ \Typ{then} ~ \neg s :: \Splice{\DV{\pi^\#}} ~ \Typ{else} ~ s :: \Splice{\DV{\pi^\#}} ~ \Typ{in} \\
       & && \magic \qspace \Typ{if} ~ v \equiv 0 ~ \Typ{then} ~ \Splicee{\SConcdeno{\SV{\es_1}}(\DV{\memo})(\Quo{\sigma^\#_{\mathit{arg}}}, \DV{\rho^\#}, \Quo{\pi^\#_1}, \Quo{\kappa^\#_1}, \Quo{\kappa^\#_1} :: \SV{\theta^\#})} ~ \\
       & && \hspace{-18.75em} \qspace \Typ{else} ~ \Splicee{\SConcdeno{\SV{\es_2}}(\DV{\memo})(\Quo{\sigma^\#_{\mathit{arg}}}, \DV{\rho^\#}, \Quo{\pi^\#_1}, \Quo{\kappa^\#_1}, \Quo{\kappa^\#_1} :: \SV{\theta^\#})}
        } \\
    & \SConcdeno{\Inst{br} ~ \SV{\ell} :: \SV{\mathit{rest}}}(\DV{\memo})(\DV{\sigma^\#}, \DV{\rho^\#}, \DV{\pi^\#}, \DV{\kappa^\#}, \DV{\theta^\#}) && = \Quote{\Splicee{\SV{\theta^\#}(\SV{\ell})}(\Splice{\DV{\sigma^\#}}, \Splice{\DV{\rho^\#}}, \Splice{\DV{\pi^\#}})} \\
    & \SConcdeno{\Inst{call} ~ \SV{x} :: \SV{\mathit{rest}}}(\DV{\memo})(\DV{\sigma^\#}, \DV{\rho^\#}, \DV{\pi^\#}, \DV{\kappa^\#}, \DV{\theta^\#}) && = \\
       & && \magic \Typ{let} ~ \{ \Typ{type}: t^m \to t^n, \Typ{locals}: \SV{\mathit{ts}}, \Typ{body}: \SV{\es} \} \coloneqq \Typ{lookupFunc}(\SV{x}) ~ \Typ{in} \\
       & && \magic \Quote{\Typ{let} ~ f \coloneqq \Splice{\DV{\memo}}(\Splicee{\lift(\SV{x})}) ~ \Typ{in} \\
       & && \magic \qspace \Typ{let} ~ \rho^\#_1 \coloneqq \mathsf{buildEnv}(\Splice{\DV{\sigma^\#}.\take_m}, \Splicee{\lift(\SV{\mathit{ts}})}) \\
       & && \magic \qspace\Typ{let} ~ \kappa^\#_1 \coloneqq \lambda (\sigma^\#_1, \rho^\#_1, \pi^\#_1). \Splicee{\SConcdeno{\SV{\mathit{rest}}}(\DV{\memo})(\Quote{\truncate{\sigma^\#_1}{n} \mathbin{+\!\!+} \Splice{\DV{\sigma^\#}}.\pop_m}, \DV{\rho^\#_1}, \DV{\pi^\#_1}, \DV{\kappa^\#}, \SV{\theta^\#})} ~ \Typ{in} \\
       & && \magic \qspace\Splice{\DV{f}}(\rho^\#_1, \Splice{\DV{\pi^\#}}, \kappa^\#_1, \Splice{\DV{\memo}}) }
  \end{alignat*}
  \vspace{-2em}
  \caption{Staged concolic semantics of \lang. The complete semantics is in
    \Cref{sec:appendix:full-definitions}.}
  \label{fig:staging:semantics}
  \vspace{-1.5em}
\end{figure}

\subsubsection*{Staging Semantic Domains}
The domains of the staged semantics are derived from the concolic semantics
(\Cref{fig:concolic:semantics}) by analyzing whether values in each domain
depend on the static program or dynamic input.
If they depend on dynamic input, we lift their type to the second stage by
wrapping it in $\mathsf{Expr}$.
The stack, environment, and path condition contain runtime values, so we lift
their types to the second stage.
Since the generated code invokes continuations with these runtime states,
continuations are also lifted.
The trail is a list of continuations, but only the continuations within the list
are residualized:
\begin{alignat*}{3}
  \theta^\# & \in \Typ{Trail^\#} && = \HL{\Typ{List[\QuoteT{Cont^\#}]}}
\end{alignat*}
A careful (manual) binding-time analysis of the concolic semantics shows that the
static source-program structure and control flow, such as $\Inst{br}~i$,
completely determine the trail structure and its access indices at compile
time. Because its structure is known statically, the list remains a
current-stage value and need not be lifted to $\Typ{Expr[List[Cont^\#]]}$.

\subsubsection*{Compiling Mutually Recursive Functions}

Since WebAssembly function definitions can be mutually recursive, directly
unrolling recursive calls could lead to non-terminating specialization.
To ensure termination in the staged semantics, we introduce a run-time
construct $\Typ{Memo}$ that memoizes compiled Wasm function definitions.
Instead of unrolling calls, we generate code that, at runtime, looks up the
callee's compiled function in $\Typ{Memo}$ and invokes it.
\begin{alignat*}{3}
    & \memo \in \Typ{Memo} &= \Typ{Map}[\Typ{Identifier}, (\Typ{Env^\#}, \Typ{PC^\#}, \Typ{Cont^\#}, \Typ{Memo}) \rightarrow \Ans]
\end{alignat*}
We define a module semantics $\MConcdeno{\cdot}$ (\Cref{fig:staging:module})
that traverses all functions in a Wasm module only once and produces the
$\Typ{Memo}$ with their staged function definitions.
$\Typ{Memo}$ is self-referential: the functions in $\Typ{Memo}$ access
the $\Typ{Memo}$ itself to call to other functions.
When later interpreting a $\Inst{call}$ instruction, we generate code
looking up the corresponding compiled function in the memo.
In practice, we implement staging with LMS \cite{DBLP:conf/gpce/RompfO10}, which
internally memoizes staged function definitions, so recursive references reuse
an existing definition rather than recompile its body. 
This prevents recursive calls from causing non-terminating specialization
and eliminates the need to maintain $\Typ{Memo}$ explicitly.

\subsubsection*{Adding Stage Annotations to Concolic Semantics}

Now, we can refactor the concolic semantics into its staged version.
The staging annotations indicate the evaluation time of expressions, staged
expressions marked with $\Quote{\cdot}$ are delayed to runtime, while those
unmarked are evaluated during specialization.
Compared to the concolic semantics $\Concdeno{\cdot}$, we replace the pattern
matching on the stack $\sigma^\#$ with explicit stack manipulation operations
$\Inst{head}$ and $\Inst{tail}$ for clarity.
$\SConcdeno{\cdot}$ does not evaluate the final result of the program, instead, it
constructs next-stage expressions that compute the final result when
executed. In the following, we explain a few representative cases.

\begin{itemize}[leftmargin=1.75em]
  \item In the case of $t.\Inst{const} ~ \SV{c}$, we splice the current stack
    $\sigma^\#$ into a staged expression that constructs the new stack.
    Then, we call $\SConcdeno{\cdot}$ recursively with the staged new stack to build
    the expression that computes the final result.
    Since $\SV{c}$ is a current stage value, we call meta-function $\lift$ to convert it to a
    next stage value before using it in the staged expression.
  \item For blocks, $\SConcdeno{\Inst{block} \dots}$ produces a staged let-expression 
    that defines the continuation $\kappa^\#_1$ and splices the result
    of applying $\SConcdeno{\SV{es}}$ to the block body.

    We define continuation $\kappa^\#_1$ by splicing the call of $\SConcdeno{\SV{rest}}$.
    The continuation $\kappa^\#_1$ maintains a valid stack matching the return
    type of this $\Inst{block}$ when calling $\SConcdeno{\SV{rest}}$:
    $\kappa^\#_1$ first truncates the top $n$ elements from the incoming stack $\sigma^\#_1$ and appends
    the remaining stack $\DV{\sigma^\#}$ after dropping its top $m$ elements.
    When invoked, $\kappa^\#_1$ receives the latest environment $\rho^\#_1$ and
    path condition $\pi^\#_1$. 
    Since $\rho^\#_1$ and $\pi^\#_1$ are variables bound by a staged
    $\lambda$-term, we quote them when referencing them in the spliced
    expression.

    Finally, to generate code computing the $\Inst{block}$ result,
    we call $\SConcdeno{\SV{es}}$ on the block body using the top $m$
    elements of the stack $\DV{\sigma^\#}$, the continuation $\Quo{\kappa^\#_1}$,
    and the trail $\Quo{\kappa^\#_1} :: \SV{\theta^\#}$.

  \item The $\Inst{loop}$ case is similar to $\Inst{block}$, except that we
    additionally construct a staged continuation $\kappa^\#_2$ representing the
    start of the loop body. Invoking $\kappa^\#_2$ with the current stack,
    environment, and path conditions produces the final result of this case.

  \item For the $\Inst{if}$ instruction, we first pop the condition $\la v, s\ra$
    from the stack and prepare the block arguments by taking
    the top $m$ items from the remaining stack.
    The remaining computation after the $\Inst{if}$ is residualized into
    continuation $\kappa^\#_1$.
    We build a new path condition $\pi^\#_1$ by prepending $s$ for
    the true branch and $\neg s$ for the false branch to the current path
    condition $\pi^\#$.

    Finally, we generate a staged conditional expression
    whose branches splice the current-stage calls $\SConcdeno{\SV{es_1}}$
    and $\SConcdeno{\SV{es_2}}$, respectively.
    We quote $\Quo{\sigma^\#_{\mathit{arg}}}$, $\Quo{\pi^\#_1}$, and
    $\Quo{\kappa^\#_1}$ when passing them to $\SConcdeno{\cdot}$, since these
    variables are created inside the staged expression, while we pass
    $\DV{\rho^\#}$ directly because it is a current-stage variable bound to a
    staged expression.

  \item In the case of $\Inst{call}$, we first retrieve the function identifier
    using the compile-time function $\Inst{lookupFunc}$.
    With the identifier $x$, we look up the run-time $\Typ{Memo}$-table $\memo$ to retrieve
    the compiled function $f$.
    We prepare the argument environment $\rho^\#_1$ and continuation $\kappa^\#_1$,
    similar to the unstaged counterpart.
    Finally, we generate a runtime call to $f$ with the prepared arguments.
\end{itemize}

\subsubsection*{A Working Example}

\newcommand{\lvone}{\hspace{2em}}
\newcommand{\lvtwo}{\hspace{4em}}

\begin{figure}[t]\small
  \begin{subfigure}[t]{0.43\textwidth}
  \begin{lstlisting}[language=Wasm, numbers=none]
    local.get 0
    local.get 1
    i32.gt_s
    (if (result i32)
      (then local.get 0 
            br 0)
      (else local.get 1
            br 0))
  \end{lstlisting}
  \vspace{-0.5em}
  \caption{The input Wasm program computing max.}\label{fig:staging:input}
  \end{subfigure}
  \hfill
  \begin{subfigure}[t]{0.55\textwidth}
  \begin{lstlisting}[breaklines=false]
    // current env $\rho$, path cond $\pi$, continuation $\kappa$
    let (v1, s1) = $\rho$(0) in
    let (v2, s2) = $\rho$(1) in
    let (v3, s3) = (v1 > v2, s1 >$^\#$ s2) in
    // $\kappa_1$ represents the continuation after the if
    let $\kappa_1$ = $\lambda$.($\sigma_1$, $\rho_1$, $\pi_1$). $\dots$ in
    if v3 == 0 then  $\kappa_1$($\rho$(0) :: [], $\rho$, s3 :: π)
               else  $\kappa_1$($\rho$(1) :: [], $\rho$, $\neg$s3 :: π)
  \end{lstlisting}
  \vspace{-0.5em}
  \caption{The compiled concolic execution program.}\label{fig:staging:output}
  \end{subfigure}
  \vspace{-1.2em}
  \caption{An example of compilation via staging.}
  \label{fig:staging:wasm}
  \vspace{-1.5em}
\end{figure}

\Cref{fig:staging:input} shows a Wasm block that computes the maximum of two
integers stored in local variables using an $\Inst{if}$ instruction.
Its compiled concolic execution, obtained by applying the staged concolic semantics $\SConcdeno{\cdot}$,
is shown in \Cref{fig:staging:output}.
Since the dispatching on the instructions has been performed during the staging, the
specialized program contains only the computations necessary to compute the
maximum of two integers and build the path conditions. The administrative
instructions such as stack resizing and pushing/popping frames have been compiled away.
In other words, the interpretation overhead has been eliminated.

\subsection{Staged Concolic Scheduler} \label{sec:compilation:scheduler}

We adapt the exploration scheduler to work with the compiled concolic program and show it
in \Cref{sec:appendix:full-definitions}.
The scheduler is a runtime function that invokes the staged
module semantics $\MConcdeno{\mathit{mod}}$ to compile all Wasm functions. 
Then, using MSP's standard facility $\run$, the scheduler evaluates the generated
staged expression to obtain a map of compiled functions.

The compiled program executes only the essential concolic logic for the source
program, eliminating interpretation overhead from instruction dispatching. In
practice, the scheduler can be implemented in the same language as the generated
program and invoke the compiled code directly. For example, in \tool
implementation (\Cref{sec:impl}), we generate C++ code for concolic execution of
Wasm programs, and the scheduler is also implemented in C++ as part of the runtime,
which will be linked with the generated code.

\subsection{Termination and Correctness of Compilation}

Since we are building a compiler, we must ensure that our staging process
terminates and produces
staged programs that preserve the source semantics.
The staged concolic interpreter $\SConcdeno{\cdot}$ avoids diverging for
recursive Wasm function calls by using the $\Typ{Memo}$ to store
staged function definitions.
We construct a staged function call by looking up its definition from
$\Typ{Memo}$, rather than directly evaluating (\ie, specializing) the
function's body.

Compared to building a concolic-compiler from scratch, our approach is not only
simpler but also easier to reason about its correctness, since the
unstaged interpretation serves as the specification:
$$
\forall \mathit{input}, \mathit{fs}, \es. \run(\SConcdeno{\es}(\MConcdeno{\mathit{fs}}))(\mathit{input}, \rho^\#_0,  \pi^\#_0, \kappa_{\text{halt}}, [\kappa_{\text{halt}}]) \equiv \Concdeno{\es} (\mathit{input}, \rho^\#_0,  \pi^\#_0, \kappa_{\text{halt}},[\kappa_{\text{halt}}])
$$
The equation states that the staging semantics $\SConcdeno{\cdot}$
and $\MConcdeno{\cdot}$ must agree with the original concolic semantics
$\Concdeno{\cdot}$.
This property requires the source program and the specialized program always
return the same result when given the same $\mathit{input}$ and other initial
arguments, including empty environment $\rho^\#_0$, empty path condition
$\pi^\#_0$, halt continuation $\kappa_{\text{halt}}$, and trail with only $\kappa_{\text{halt}}$.
The specialized program is obtained by applying $\run$ on the staged term
produced by $\SConcdeno{\cdot}$ and $\MConcdeno{\cdot}$.
On the right-hand side of the equation, $\Concdeno{\cdot}$ does not take $\mathit{fs}$
as input, because it implicitly accesses function definitions in the module via
$\mathsf{lookupFunc}$.

\section{Snapshotting Compiled Concolic Execution with Continuations}
\label{sec:timetravel}

The scheduler defined in \Cref{sec:compilation:scheduler} re-executes the
program from the beginning for each new input, causing redundant work along
shared path prefixes. This redundancy becomes costly for programs with deep
execution paths.
We now present the snapshot-reuse optimization, which uses continuations to
eliminate redundant execution in the compiled concolic program and its
scheduler.

Conceptually, concolic execution explores an implicit execution
tree whose nodes represent branch points, leaves mark execution ends, and
root-to-leaf paths correspond to complete traces. As the tree expands, both the
number of leaves and the re-execution cost grow exponentially.
The key idea is to save both program states \emph{and} continuations at
branch points, which together we call \emph{snapshots}.
Resuming from snapshots explores new paths instead of re-executing shared path segments.
\looseness=-1

\subsubsection*{From State-Passing to Global Mutable States}
Resuming from a snapshot requires symbolic states sufficient to reconstruct the
entire program state.
However, the staging semantics in \Cref{fig:staging:semantics} works on a local stack and
environment that are passed as parameters.
These parameters capture only a portion of the program state rather than the entire state.
For example, in \Cref{fig:staging:semantics}, we always ensure the function's local
stack is initialized to empty before entering the function body, hence the
caller's stack is not visible to the callee.
Therefore, we refactor the state-passing interpreter to use mutable global
variables, \ie, $\DV{\sigma_g^\#}$, $\DV{\rho_g^\#}$, and $\DV{\pi_g^\#}$, which
hold references to the \emph{entire} stack, environment, and path condition,
respectively.
Accordingly, the continuation function type does not take any argument and
performs effects by updating these globals:
$\kappa^\# \in \Typ{Cont^\#} = () \to \Typ{Ans}$.

\subsubsection*{Capturing Snapshots}
A snapshot is a triple consisting of continuations, symbolic stacks, and symbolic
environments. Together they capture necessary information to resume execution from a
branch point.
We also redefine $\Typ{PC^\#}$ to include snapshots of the \emph{untaken} branch
in addition to the constraints of the \emph{taken} branch:
\begin{alignat*}{3}
  \Sigma \in \Typ{Snapshot}   & = \Typ{Cont^\#} \times \Typ{List[Symbolic]} \times \Typ{List[Symbolic]} \\
  \pi^\# \in \Typ{PC^\#} & = \Typ{List[Symbolic \times \Typ{Snapshot}]}.
  \end{alignat*}

\renewcommand{\magic}{\hspace{-16em}}
\newcommand{\shift}{\ensuremath{\mathsf{shift}}}
\newcommand{\length}{\ensuremath{\mathsf{length}}}
\newcommand{\tru}{\ensuremath{\mathit{tru}}}
\newcommand{\fls}{\ensuremath{\mathit{fls}}}

\begin{figure}\small
\judgement{Staged Snapshotting Concolic Continuation Semantics}{}

\vspace{.75em}
Global stack $\DV{\sigma_g^\#}$, global environment $\DV{\rho_g^\#}$, global path condition $\DV{\pi_g^\#}$
  \begin{alignat*}{3}
    & \SConcdeno{\cdot} : \mathrlap{\Typ{List[Inst]} \rightarrow (\HL{\QuoteT{\Typ{Cont^\#}}} \times \HL{\Typ{Trail^\#}}) \rightarrow \HL{\QuoteT{\Typ{Ans}}}} \\
    & \SConcdeno{\Inst{if} ~ (t^m \to t^n) ~ \SV{\es_1} ~ \SV{\es_2} :: \SV{\mathit{rest}}}(\DV{\pi^\#}, \DV{\kappa^\#}, \SV{\theta^\#}) && = \\
       & && \magic \Quote{ \Typ{let} ~ \la v, s\ra \coloneqq \Splice{\DV{\sigma_g^\#}}.\pop^! ~ \Typ{in} \\
       & && \magic \qspace \Typ{let} ~ (\_ \pconcat{m} \sigma^\#_{\mathit{old}}) \coloneqq \Splice{\DV{\sigma_g^\#}} ~ \Typ{in} \\
       & && \magic \qspace \Typ{let} ~ \kappa^\#_1 \coloneqq \lambda (). \Splice{\DV{\sigma_g^\#}} \leftarrow \truncate{\Splice{\DV{\sigma_g^\#}}}{n} \mathbin{+\!\!+} \sigma^\#_{\mathit{old}};\ \Splicee{\SConcdeno{\SV{\mathit{rest}}}(\DV{\kappa^\#}, \SV{\theta^\#})} ~ \Typ{in} \\
       & && \magic \qspace \Typ{let} ~ \kappa^\#_{\tru} \coloneqq \lambda (). \Splicee{\SConcdeno{\SV{\es_1}}(\Quo{\kappa^\#_1}, \Quo{\kappa^\#_1} :: \SV{\theta^\#})} ~ \Typ{in} \\
       & && \magic \qspace \Typ{let} ~ \kappa^\#_{\fls} \coloneqq \lambda (). \Splicee{\SConcdeno{\SV{\es_2}}(\Quo{\kappa^\#_1}, \Quo{\kappa^\#_1} :: \SV{\theta^\#})} ~ \Typ{in} \\
       & && \magic \qspace \Typ{let} ~ \Sigma_{\tru} \coloneqq \la\kappa^\#_{\tru}, \Splice{\DV{\sigma_g^\#}}.\mathsf{sym}, \Splice{\DV{\rho_g^\#}}.\mathsf{sym}\ra ~ \Typ{in} \\
       & && \magic \qspace \Typ{let} ~ \Sigma_{\fls} \coloneqq \la\kappa^\#_{\fls}, \Splice{\DV{\sigma_g^\#}}.\mathsf{sym}, \Splice{\DV{\rho_g^\#}}.\mathsf{sym}\ra ~ \Typ{in} \\
       & && \magic \qspace \Typ{if} ~ v \equiv 0 ~
              \Typ{then} ~ \Splice{\DV{\pi^\#}}.\mathsf{push}^!(\la\neg s, \Sigma_{\mathit{tru}}\ra);\ \kappa^\#_{\mathit{fls}}() ~
              \Typ{else} ~ \Splice{\DV{\pi^\#}}.\mathsf{push}^!(\la s, \Sigma_{\mathit{fls}}\ra);\ \kappa^\#_{\mathit{tru}}() }
  \end{alignat*}
  \vspace{-2em}
  \caption{The staged snapshotting concolic semantics of \lang.}
  \label{fig:snapshot:semantics}
  \vspace{-1.5em}
\end{figure}

\Cref{fig:snapshot:semantics} highlights the modification to the staged
semantics in \Cref{fig:staging:semantics} for handling \Inst{if} instructions.
The key change is to represent the computations of the ``then'' and ``else'' branches as
next-stage continuations, $\kappa^\#_{\mathit{tru}}$ and
$\kappa^\#_{\mathit{fls}}$.
Using these continuations and the current symbolic states, we create the corresponding
snapshots $\Sigma_{\mathit{tru}}$ and $\Sigma_{\mathit{fls}}$.
Since the stack consists of concolic pairs, we extract the symbolic stack using
$\sigma^\#.\mathsf{sym}$, and apply the same to the environment. We then update
the global path condition according to the branch taken and invoke the
corresponding continuation.

Since now the symbolic stack/environment are global mutable states,
operations on them also become imperative.
At the beginning, we pop the concolic condition from the global stack
$\DV{\sigma^\#}$ by $\pop^!$.
We mark these effectful operations on global states with a superscript exclamation mark.
Recall that in WebAssembly, the type of $\Inst{if}$ regulates the input/output
stack: the branches can access top $m$ elements on the stack, and each branch
leaves the stack with $n$ new elements after removing the $m$ inputs.
Therefore, we save the stack below those $m$ inputs into $\sigma^\#_{\mathit{old}}$.
When resuming via $\kappa^\#_1$, we use the result of prepending the $n$ outputs
to $\sigma^\#_{\mathit{old}}$ to restore the global stack $\DV{\sigma_g^\#}$.

\subsubsection*{Resuming from Snapshots}

\begingroup
\begin{algorithm}[t]
  \small
  \caption{Concolic Snapshot Scheduler}
  \label{fig:snapshot:driver}
  \begin{algorithmic}[1]
    \Function{\textsc{Scheduler}}{$mod \in \Typ{Module}$, $f \in \Typ{Identifier}$, $\mathit{input} \in \Typ{List[Symbolic]}$}
    \State $\mathit{memo} \gets \mathsf{run} ~ \MConcdeno{mod}$
    \State $\mathit{entry} \gets \mathit{memo}(f)$
    \State $WL \gets \HL{\{\, \la [], \la \lambda () . \mathit{entry}(\kappa^\#_{halt}, \mathit{memo}), \mathit{input}, []\ra\ra \, \}}$ \Comment{$\HL{WL \in \Pow{\Typ{List[Symbolic]} \times \Typ{Snapshot}}}$}
    \State $Seen \gets \varnothing$ \Comment{$\HL{Seen \in \Pow{\Typ{List[Symbolic]}}}$}
      \While{$WL \neq \varnothing$}
        \State pop $\la\pi^\#, \HL{\la\kappa, \sigma^\#, \rho^\#\ra}\ra$ from $WL$
        \IIf{$\pi^\# \in Seen$} \textbf{continue}
        \State $m \gets \mathit{solve}(\pi^\#)$
        \IIf{$m = \UNSAT$} \textbf{continue}
        \State \begin{tabular}[t]{@{}l@{\hspace{0.5em}}l@{}}
          $\la \sigma_g^\#, \rho_g^\#, \pi_g^\# \ra \gets \la \mathit{instantiate}(\sigma^\#, m), \mathit{instantiate}(\rho^\#, m), \pi^\# \ra$
          & \hspace{-0.2em}$\triangleright$ instantiate global stack/environment \\
          & \hspace{-0.2em}$\triangleright$ update global path condition
        \end{tabular}
        \State $\HL{\Pi^\# \gets \kappa()}$
        \State $Seen \gets Seen \cup \{\HL{\Pi^\#.\mathsf{pc}}\}$ \Comment{$\Pi^\#.\mathsf{pc}$ extracts a list of symbolic conditions}
        \ForAll{$\HL{\la\pi_1, \Sigma_1\ra}\in \textsc{Negate}(\Pi^\#)$}
          \State $WL \gets WL \cup \{\HL{\la\pi_1, \Sigma_1\ra}\}$
        \EndFor
      \EndWhile
    \EndFunction

    \Function{\textsc{Negate}}{$\Pi^\# \in \Typ{PC^\#}$} $\rightarrow \HL{\Typ{List}[\Typ{List[Symbolic]} \times \Typ{Snapshot}]}$ \dots  \Comment{complete definition in the appendix}
    \EndFunction
  \end{algorithmic}
\end{algorithm}
\endgroup

Algorithm 2 shows a modified scheduler that reconstructs concrete states from
snapshots to resume the execution without re-executing from the beginning.

The worklist pairs path conditions and their corresponding snapshots.
It is initialized with an empty path condition and a snapshot of the program entry.
In each iteration, the scheduler pops a pair and checks the
condition's feasibility as in \Cref{sec:concolic}.
If the condition is satisfiable, it uses the model to instantiate
the snapshot's symbolic states to update the global states.
The scheduler then invokes the continuation stored in the snapshot to resume
execution from the branch point.
The execution returns a new $\Pi^\#$ pairing taken-branch path conditions with
snapshots of the untaken branches. 
Finally, the scheduler calls $\textsc{Negate}$ to append
the untaken branches' path conditions and snapshots to the worklist for future
exploration.
Thus, continuation-based resumption avoids re-executing shared path
prefixes while retaining the performance of compiled execution.

\subsubsection*{Improving Asymptotic Complexity}
By applying snapshots to avoid re-execution, we improve the asymptotic complexity of
concolic execution.
Suppose the implicit execution tree of a concolic execution contains $n$ nodes.
The tree's height ranges from $O(\log n)$ (balanced) to $O(n)$ (unbalanced).
The complexity of exploring a path from root to leaf takes between $O(\log n)$ and
$O(n)$, and exploring all paths from the root takes between $O(n\log n)$ to
$O(n^2)$.
With snapshots, however, each exploration starts from a fresh node that has
never been visited before.
As a result, every node is visited exactly once, reducing the overall
complexity to $O(n)$ in the worst case.

\section{Deciding When to Use Snapshots} \label{sec:heuristic}

\Cref{sec:timetravel} introduced snapshots to uniformly avoid re-executing
previously explored program prefixes.
Its practical effectiveness, however, depends on both the depth of the shared
execution prefix and the size of the resumed symbolic state.
For programs whose execution paths share only shallow prefixes, creating and instantiating
snapshots may cost more than re-executing the program.
Thus, applying snapshot reuse uniformly, as in \Cref{fig:snapshot:driver}, may
yield little benefit or even degrade performance for such programs.

\begin{wrapfigure}[8]{r}[0.4em]{0.23\columnwidth}
  \centering
  \vspace{-1.2em}
  \includegraphics[scale=0.88]{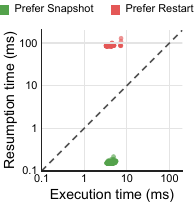}
  \vspace{-1em}
    \end{wrapfigure}
The figure on the right illustrates this trade-off for the quicksort benchmark
used in our evaluation (\Cref{sec:rq2}).
The benchmark contains 862 complete paths and 861 branch points.
Each of the 861 plotted points represents the choice at one branch point between
resuming from its snapshot and re-executing from the program entry to reach it.
The x-axis shows the re-execution cost and the y-axis the snapshot-resumption
cost.
Of these 861 choices, 283 lie above the dashed equal-cost line, indicating that
snapshot resumption is more expensive than re-execution.
In this example, reusing snapshots uniformly at these unprofitable points would
therefore make the overall execution slower than uniform re-execution,
as the overall cost of these 283 snapshots is substantially higher than the cost
of re-execution.

Additionally, the snapshot mechanism in \Cref{fig:snapshot:semantics} captures only
the states represented in the semantics.
However, realistic programs interact with external environments
(\eg, the OS), whose states are typically not preserved in a snapshot.
When an execution path involves such interactions, the symbolic state recorded
in the snapshot may no longer be reachable in a concrete execution.
In that case, resuming from the snapshot may break the soundness of concolic
execution.

To make snapshot reuse both effective and sound, we need a strategy for deciding
when to resume execution from a snapshot and when re-execution from the
beginning is preferable.
This section develops a heuristic for making this choice based on cost
estimation while enforcing soundness.

\subsubsection*{Cost of Re-execution vs. Instantiation}
The main cost of instantiating a snapshot comes from reconstructing the
concrete state from the snapshot, which involves traversing all symbolic
expressions in the symbolic stack and environment and evaluating them with
respect to the SMT model.
We estimate this cost by the number of symbolic expressions and denote it
as $C_{\text{Ins}}$:
\begin{alignat*}{3}
  C_{\text{Ins}} & = \delta * (\mathsf{sizeOfSymExp}(\sigma^\#) + \mathsf{sizeOfSymExp}(\rho^\#))
\end{alignat*}
where parameter $\delta$ is the average timing cost of evaluating symbolic expressions in the solver.
In practice, we can obtain $\delta$ by measuring the evaluation time of a large set of symbolic
expressions.
During runtime, we also record the execution time from the beginning of the
program to the branch point as the re-execution cost, denoted $C_{\text{Re}}$.

\subsubsection*{Soundly Handling External Calls}
External calls invoke functions whose implementations are unavailable to the
concolic or symbolic execution engine, such as system calls or library calls.
One way to handle external calls is to execute them concretely and treat the
returned concrete values as their symbolic counterparts in subsequent concolic
execution.

However, this strategy can make snapshot reuse unsound: a snapshot based on an
observed concrete return value assumes that the external call returns the same
value across executions.
In a fresh execution from the beginning, the call may return a different value
because its arguments or the external environment have changed.
As a result, the symbolic states captured by the snapshot may be invalid,
and the resumed execution may fail to correspond to a feasible program path.

To avoid creating unsound snapshots, we can conservatively disable resumption of
snapshots whose execution prefix contains an external call.
For such snapshots, instead of computing $C_{\text{Ins}}$ using
$\mathsf{sizeOfSymExp}$, we set it to infinity, ensuring that the scheduler
re-executes the program from beginning.
A more sophisticated concolic execution engine can handle external calls soundly
by symbolically modeling the semantics of external functions.
We leave the integration of sound external call handling and its interaction
with snapshot reuse as future work.

\subsubsection*{Heuristic Decision}

The decision of whether to resume from a snapshot at a branch therefore reduces
to comparing the cost of instantiation with that of re-execution.
At runtime, we compute the ratio $C_{\text{Re}} / C_{\text{Ins}}$.
If this ratio exceeds 1, re-execution is deemed more expensive than
instantiation, and we therefore prefer to resume from the snapshot.
Otherwise, we re-execute the program from the beginning.
To implement this heuristic, we extend \Typ{Snapshot} with an additional field
\Typ{Cost} = \Typ{Int}, which estimates the cost of reaching the program point
at which the snapshot was created:
\begingroup
\setlength{\abovedisplayskip}{4pt}
\setlength{\belowdisplayskip}{4pt}
\[
  s \in \Typ{Snapshot}
  = \Typ{Cont^\#} \times \Typ{List[Symbolic]}
  \times \Typ{List[Symbolic]} \times \HL{\Typ{Cost}}.
\]
\endgroup
\Cref{sec:appendix:full-definitions} presents the heuristic scheduler.
In each iteration, the scheduler compares the cost stored in the snapshot
against the estimated cost of instantiation, represented by the size of the
symbolic state.

\section{Implementation} \label{sec:impl}
We scale our approach to a substantially larger subset of WebAssembly and
implement it in \tool, a compiler written in Scala using the
Lightweight Modular Staging (LMS) framework \cite{DBLP:conf/gpce/RompfO10}.
\looseness=-1

\subsection{Compilation via Staging}

At the core of \tool is a staged concolic interpreter for WebAssembly,
derived from our staged concolic semantics with snapshots.
Compilation is performed by specializing this interpreter to a given Wasm program.
Unlike the quasi-quotation and splicing presentations in earlier sections,
the implementation uses LMS \cite{DBLP:conf/gpce/RompfO10}, which offers convenient
staging APIs through operator overloading.
As a result, while the implementation follows the same staging semantics and
type-level stage discipline, it does not use explicit quasi-quotation/splicing syntax.
LMS also provides a graph-based IR with built-in optimizations
\cite{DBLP:journals/pacmpl/BracevacWJAJBR23}, as well as multiple code-generation
backends, including C++.
The core staged interpreter of \tool comprises 1.5K lines of Scala code.

\tool generates efficient C++ code for concolic execution in CPS.
Each WebAssembly function is compiled into a C++ function that takes an additional
continuation argument.
Following the refactoring described in \Cref{sec:timetravel}, the stack, environment,
and path condition are maintained as global mutable runtime states, and the generated
code operates directly on them.
Continuations therefore do not need to thread these states explicitly.
Moreover, the continuation trail disappears entirely in the generated code,
since all computation over the trail is resolved at the first stage (\Cref{sec:ce-compiler}).
WebAssembly stack and frame operations compile directly to operations on the corresponding
global runtime structures.
To support efficient state snapshotting, we represent the stack and environment
using persistent data structures from Immer
\cite{DBLP:journals/pacmpl/Puente17}, which provide efficient copy and update
operations while reducing the need to copy large amounts of memory.

\subsection{Supporting More WebAssembly Features}
\tool supports all numeric types in Wasm 2.0 \cite{wasmspec2}, including
32-/64-bit integers and floating-point numbers.
Symbolically, these values are represented using the bit-vector and floating-point theories
of \Zthree \cite{DBLP:conf/tacas/MouraB08}.
Beyond the features covered in our core formalization, \tool supports a large
fragment of commonly used WebAssembly features, including additional arithmetic operations,
function tables, indirect calls, global variables, and linear memory.
Supporting these additional features ensures that \tool can be applied to real-world Wasm programs.
The implementation of these features still follows the semantics-based approach,
and here we discuss the implementation of some of these features,
including global variables, linear memory, and indirect control flow.

\subsubsection*{Global Variables}
As described in \Cref{sec:timetravel}, we have refactored the semantics by lifting
the stack, environment, and path condition into mutable global states.
To support WebAssembly global variables, we add a separate global component to the
semantics.
In \tool, WebAssembly globals are compiled into a C++ global data structure,
and global-variable instructions (\eg, \code{global.get} and \code{global.set})
are translated into direct operations on it.
As with the stack and environment, we use persistent data structures from Immer
\cite{DBLP:journals/pacmpl/Puente17} to represent global state, enabling efficient snapshotting.

\subsubsection*{Linear Memory}
Support for linear memory is essential for symbolic reasoning about realistic
data structures, which are typically allocated in WebAssembly memory.
In the official WebAssembly semantics, linear memory is modeled as a mapping
from addresses to bytes.
For concolic execution, we refine this model by representing each address as
a pair of concrete and symbolic byte values.

WebAssembly provides memory operations over multiple data types,
including 32-/64-bit integers and floating-point numbers.
To handle them uniformly, we follow \WASP and implement memory operations
as byte-level operations on both concrete and symbolic memory.
For example, \code{i32.load} reads four consecutive bytes and combines them
into a 32-bit value, while \code{i32.store} decomposes a 32-bit value into
four bytes and writes them back to consecutive addresses.
In the implementation, \tool represents memory using persistent
data structures, enabling efficient snapshotting.

\subsubsection*{Indirect Control Flow}
WebAssembly indirect control flow instructions such as \code{call_indirect}
allow the call target to be resolved at runtime rather than fixed statically.
This mechanism helps compile higher-level languages with dynamic dispatch to WebAssembly.
The WebAssembly standard assigns a unique index to each function,
and a runtime global table stores these indices.
The \code{call_indirect} instruction uses its operand to look up this
table and resolve the target.
Following the specification, \tool initializes a global table mapping WebAssembly
function indices to C++ function pointers at program startup.
This table is initialized when the compiled program starts and is used to
resolve indirect calls at runtime.
We compile each \code{call_indirect} instruction into a table lookup followed by
a call to the resolved function.

\subsubsection*{Limitations}
As a prototype implementation, \tool already supports a substantial subset of
WebAssembly 2.0 \cite{wasmspec2}, but it does not yet cover all features defined in the WebAssembly
standard.
\tool currently does not support several additional WebAssembly
features, such as SIMD instructions, function references,
and module import/export. Nevertheless, as long as the specification provides clear semantics for
these features, extending \tool to support these features would be straightforward and
requires moderate engineering effort.

\section{Empirical Evaluation} \label{sec:eval}

This section evaluates the performance of \tool with respect to the following
three questions:
\begin{enumerate}[leftmargin=3.5em, label=\textbf{RQ\arabic*:}]
  \item How much speedup does compilation provide over interpretation-based approaches?
  \item When snapshot and the heuristic are effective in \tool?
  \item Is \tool effective in finding real bugs?
\end{enumerate}

\subsubsection*{Environment and Setup}
All experiments ran on Ubuntu 22.04 LTS with an AMD Ryzen 7 7735H CPU and 16 GB
RAM. We used \Zthree 4.8.12 and compiled generated code with Clang 22.1.0 at -O3.
For the baseline, we used \WASP version 0.2.3, the latest release at submission
time of this paper.

For performance measurement, we decompose the total wall-clock time into
instruction-execution time and SMT-solving time. Our evaluation focuses on
instruction execution, reporting the arithmetic mean over 5 runs for each benchmark.
We also report end-to-end execution time (including time spent in SMT solvers) for
completeness, although it depends on symbolic-expression encodings and query
optimizations, which are beyond the scope of this paper.
To validate the generated SMT queries and ensure the fairness of the comparison,
we verify that the number of explored paths matches that of the baseline \WASP
\cite{DBLP:conf/ecoop/MarquesS0A22}.

\subsubsection*{Benchmarks}

We evaluate against four groups of benchmarks, shown in \Cref{tab:rq1-uniform}.
To assess performance on practical workloads, we adopt
two benchmarks from prior work: (1) the B-Tree benchmark from \WASP
\cite{DBLP:conf/ecoop/MarquesS0A22}, and (2) the Collections-C test suites
\cite{srdja_collections_c}, which exercise various data structures from the
Gillian project \cite{DBLP:conf/pldi/SantosMAG20}.
Compared with their original configurations in \WASP, we use larger symbolic
inputs and explore more paths for both the B-Tree benchmark
\cite{DBLP:conf/ecoop/MarquesS0A22} and the Collections-C benchmark
\cite{srdja_collections_c} to better approximate practical workloads.
The B-Tree benchmark contains 26 cases, while the Collections-C benchmark
contains 143 across 8 modules, yielding 169 cases in total.

We include two micro-benchmarks to understand the performance
characteristics of snapshot reuse: (3) an arithmetic evaluator and (4) a
quicksort implementation.
These two benchmarks are designed to isolate the
performance effects of snapshot reuse (RQ2). The quicksort benchmark is written
in C and compiled to Wasm, whereas the arithmetic evaluator is written directly
in Wasm. The arithmetic-evaluator benchmark contains 9 Wasm programs, each of
which parses and evaluates a string-represented arithmetic expression. The programs
vary in expression size, measured by the number of arithmetic operators, and
the number of explored paths.

The quicksort benchmark varies both the input-array size and the number of
symbolic elements. We initialize the symbolic inputs with either atomic symbolic
variables or complex symbolic arithmetic expressions, allowing us to study how
symbolic-state complexity affects snapshot reuse.

To answer RQ3, we use two Collections-C benchmarks from \WASP
\cite{DBLP:conf/ecoop/MarquesS0A22} to evaluate bug-detection effectiveness.
Each benchmark has two versions: a \emph{buggy version}
containing two known bugs and a \emph{correct version} in which both bugs are
fixed.

\subsubsection*{Baselines}

There are several recent interpretation-based symbolic execution engines for
WebAssembly, including \Manticore \cite{DBLP:conf/kbse/MossbergMHGGFBD19},
\SeeWasm \cite{DBLP:conf/issta/HeZGWP0WC024}, \WASP
\cite{DBLP:conf/ecoop/MarquesS0A22} and \WASP's successor \Owi
\cite{DBLP:journals/corr/abs-2412-06391}.
For RQ1, we choose to use \WASP as the baseline for comparison, as it is the
only tool among them implementing the same concolic execution semantics as
\tool.
\Owi and \Manticore both implement a general parallel symbolic execution
semantics, which is not driven by concrete execution.
\SeeWasm constructs a control-flow graph for Wasm program, then symbolically
explores the CFG using BFS or DFS strategy, without concrete inputs.

However, as these tools realize similar testing capabilities for end-users,
their relative performance is worth noting to better understand where \tool
would stand.
On the same B-Tree benchmark, \Owi with 1 worker reports similar performance
comparable to \WASP ($0.8\times$ slowdown), while \Owi with 24 workers
achieves an average $4.1\times$ speedup over \WASP
\cite{DBLP:journals/corr/abs-2412-06391}.
In addition, \WASP reports $5.4\times$--$26.2\times$ speedups over \Manticore
\cite{DBLP:conf/ecoop/MarquesS0A22},
whereas \SeeWasm reports $1.2\times$--$4.8\times$ speedups over \Manticore
\cite{DBLP:conf/issta/HeZGWP0WC024}.
Taken together, these published results from the literature suggest that \WASP
is a competitive baseline for our evaluation.

To the best of our knowledge, there currently exists no instrumentation-based
concolic engine for Wasm, so we could not include such a baseline in our
evaluation.

\begin{table*}[t]
  \newcommand{\meanstd}[2]{\makebox[3.2em][r]{#1}\,$\pm$\,\makebox[3.0em][l]{#2}}
  \centering
  \caption{Performance comparison between \tool and \WASP.}
  \vspace{-1em}
  \label{tab:rq1-uniform}
  \resizebox{\textwidth}{!}{%
  \begin{tabular}{@{}l c c !{\vrule width 0.6pt} c !{\vrule width 0.6pt} c c
    !{\vrule width 0.6pt} c c c
    !{\vrule width 0.6pt} c c c
    !{\vrule width 0.6pt} c c c@{}}
    \Xhline{1.2pt}
    & & & &
    \multicolumn{2}{c!{\vrule width 0.6pt}}{\WASP} &
    \multicolumn{9}{c}{\tool} \\
    \cline{5-15}
    & & & $n_{\mathit{paths}}$ & \meanstd{$T_{\mathit{exec}}^{\WASP}$}{$\sigma_{\mathit{exec}}^{\WASP}$} & $T_{\mathit{total}}^{\WASP}$ & \meanstd{$T_{\mathit{exec}}^{\mathit{noreuse}}$}{$\sigma_{\mathit{exec}}^{\mathit{noreuse}}$} & $T_{\mathit{total}}^{\mathit{noreuse}}$ & $S_{\mathit{staging}}$ & \meanstd{$T_{\mathit{exec}}^{\mathit{snapshot}}$}{$\sigma_{\mathit{exec}}^{\mathit{snapshot}}$} & $T_{\mathit{total}}^{\mathit{snapshot}}$ & $S_{\mathit{snapshot}}$ & \meanstd{$T_{\mathit{exec}}^{\mathit{heuristic}}$}{$\sigma_{\mathit{exec}}^{\mathit{heuristic}}$} & $T_{\mathit{total}}^{\mathit{heuristic}}$ & $S_{\mathit{heuristic}}$ \\[0.5ex]
    \hline
    $n_o$ & $n_u$ & \multicolumn{1}{c}{} & \multicolumn{10}{c}{{Extended B-Tree Benchmark}} \\
    \hline
    2 & 2 &  & 12 & \meanstd{0.2237}{0.0243} & 2.14 & \meanstd{0.0092}{0.0051} & 1.22 & $24.3\times$ & \meanstd{0.0089}{0.0051} & 1.22 & $1.0\times$ & \meanstd{0.0087}{0.0045} & 1.24 & $1.1\times$ \\
    2 & 3 &  & 60 & \meanstd{2.19}{0.0724} & 15.93 & \meanstd{0.0398}{0.0093} & 12.14 & $55.0\times$ & \meanstd{0.0363}{0.0082} & 12.24 & $1.1\times$ & \meanstd{0.0362}{0.0085} & 12.28 & $1.1\times$ \\
    3 & 2 &  & 20 & \meanstd{0.5026}{0.0180} & 4.98 & \meanstd{0.0150}{0.0051} & 3.67 & $33.5\times$ & \meanstd{0.0149}{0.0055} & 3.73 & $1.0\times$ & \meanstd{0.0143}{0.0049} & 3.63 & $1.0\times$ \\
    3 & 3 &  & 120 & \meanstd{6.46}{0.2174} & 45.26 & \meanstd{0.0880}{0.0123} & 36.20 & $73.4\times$ & \meanstd{0.0783}{0.0097} & 36.03 & $1.1\times$ & \meanstd{0.0797}{0.0101} & 35.85 & $1.1\times$ \\
    4 & 1 &  & 5 & \meanstd{0.1206}{0.0046} & 1.28 & \meanstd{0.0063}{0.0045} & 0.9061 & $19.1\times$ & \meanstd{0.0061}{0.0043} & 0.9069 & $1.0\times$ & \meanstd{0.0053}{0.0038} & 0.9058 & $1.2\times$ \\
    4 & 2 &  & 30 & \meanstd{1.09}{0.0396} & 10.44 & \meanstd{0.0219}{0.0029} & 8.01 & $49.8\times$ & \meanstd{0.0203}{0.0024} & 8.11 & $1.1\times$ & \meanstd{0.0206}{0.0036} & 8.08 & $1.1\times$ \\
    4 & 3 &  & 210 & \meanstd{16.07}{0.4950} & 109.21 & \meanstd{0.1806}{0.0213} & 87.23 & $89.0\times$ & \meanstd{0.1620}{0.0201} & 88.35 & $1.1\times$ & \meanstd{0.1612}{0.0180} & 87.78 & $1.1\times$ \\
    5 & 1 &  & 6 & \meanstd{0.1172}{0.0045} & 2.15 & \meanstd{0.0065}{0.0044} & 1.58 & $18.0\times$ & \meanstd{0.0059}{0.0038} & 1.57 & $1.1\times$ & \meanstd{0.0061}{0.0038} & 1.58 & $1.1\times$ \\
    5 & 2 &  & 42 & \meanstd{2.50}{0.0870} & 20.04 & \meanstd{0.0439}{0.0166} & 16.49 & $56.9\times$ & \meanstd{0.0386}{0.0161} & 16.15 & $1.1\times$ & \meanstd{0.0339}{0.0059} & 15.97 & $1.3\times$ \\
    5 & 3 &  & 336 & \meanstd{41.04}{1.09} & 241.80 & \meanstd{0.3590}{0.0316} & 202.25 & $114.3\times$ & \meanstd{0.2940}{0.0348} & 202.04 & $1.2\times$ & \meanstd{0.3016}{0.0400} & 204.70 & $1.2\times$ \\
    6 & 1 &  & 7 & \meanstd{0.3145}{0.0134} & 3.25 & \meanstd{0.0065}{0.0012} & 2.49 & $48.4\times$ & \meanstd{0.0059}{0.0016} & 2.49 & $1.1\times$ & \meanstd{0.0057}{0.0010} & 2.48 & $1.1\times$ \\
    6 & 2 &  & 56 & \meanstd{3.60}{0.1181} & 33.49 & \meanstd{0.0557}{0.0053} & 28.93 & $64.6\times$ & \meanstd{0.0475}{0.0042} & 28.73 & $1.2\times$ & \meanstd{0.0472}{0.0046} & 28.58 & $1.2\times$ \\
    6 & 3 &  & 504 & \meanstd{116.95}{1.98} & 493.44 & \meanstd{0.6307}{0.0869} & 418.09 & $185.4\times$ & \meanstd{0.5311}{0.0758} & 416.86 & $1.2\times$ & \meanstd{0.5241}{0.0532} & 405.87 & $1.2\times$ \\
    7 & 1 &  & 8 & \meanstd{0.3716}{0.0026} & 4.65 & \meanstd{0.0086}{0.0016} & 4.11 & $43.2\times$ & \meanstd{0.0080}{0.0019} & 4.15 & $1.1\times$ & \meanstd{0.0080}{0.0020} & 4.10 & $1.1\times$ \\
    7 & 2 &  & 72 & \meanstd{5.76}{0.0583} & 57.30 & \meanstd{0.0903}{0.0081} & 54.97 & $63.8\times$ & \meanstd{0.0767}{0.0087} & 54.82 & $1.2\times$ & \meanstd{0.0757}{0.0065} & 53.99 & $1.2\times$ \\
    7 & 3 &  & 720 & \meanstd{409.15}{1.76} & 1124.46 & \meanstd{1.18}{0.1798} & 922.24 & $346.7\times$ & \meanstd{1.01}{0.1176} & 898.69 & $1.2\times$ & \meanstd{1.01}{0.1000} & 891.70 & $1.2\times$ \\
    8 & 1 &  & 9 & \meanstd{0.6124}{0.0053} & 7.22 & \meanstd{0.0141}{0.0054} & 6.54 & $43.4\times$ & \meanstd{0.0121}{0.0042} & 6.59 & $1.2\times$ & \meanstd{0.0115}{0.0025} & 6.57 & $1.2\times$ \\
    8 & 2 &  & 90 & \meanstd{10.38}{0.0143} & 92.83 & \meanstd{0.1460}{0.0283} & 101.61 & $71.1\times$ & \meanstd{0.1279}{0.0258} & 102.27 & $1.1\times$ & \meanstd{0.1263}{0.0237} & 100.41 & $1.2\times$ \\
    9 & 1 &  & 10 & \meanstd{0.9422}{0.0103} & 10.62 & \meanstd{0.0162}{0.0027} & 10.51 & $58.2\times$ & \meanstd{0.0141}{0.0021} & 10.50 & $1.1\times$ & \meanstd{0.0137}{0.0019} & 10.49 & $1.2\times$ \\
    9 & 2 &  & 110 & \meanstd{18.16}{0.1377} & 149.14 & \meanstd{0.2056}{0.0414} & 165.65 & $88.3\times$ & \meanstd{0.1798}{0.0318} & 166.51 & $1.1\times$ & \meanstd{0.1828}{0.0403} & 168.10 & $1.1\times$ \\
    9 & 3 &  & 1320 & \meanstd{1970.68}{11.36} & 4087.60 & \meanstd{3.02}{0.5759} & 2844.54 & $652.5\times$ & \meanstd{2.36}{0.0586} & 2537.99 & $1.3\times$ & \meanstd{2.64}{0.5663} & 2946.57 & $1.1\times$ \\
    10 & 1 &  & 11 & \meanstd{0.9066}{0.0203} & 15.52 & \meanstd{0.0205}{0.0032} & 14.31 & $44.2\times$ & \meanstd{0.0174}{0.0033} & 14.22 & $1.2\times$ & \meanstd{0.0185}{0.0035} & 14.87 & $1.1\times$ \\
    10 & 2 &  & 132 & \meanstd{28.65}{0.6753} & 225.18 & \meanstd{0.2829}{0.0389} & 243.39 & $101.3\times$ & \meanstd{0.2441}{0.0338} & 244.59 & $1.2\times$ & \meanstd{0.2392}{0.0331} & 236.70 & $1.2\times$ \\
    10 & 3 &  & 1716 & - & \TO & \meanstd{4.44}{1.52} & 4333.33 & - & \meanstd{3.37}{0.0906} & 3817.26 & $1.3\times$ & \meanstd{3.65}{0.7807} & 4260.93 & $1.2\times$ \\
    11 & 1 &  & 12 & \meanstd{0.9233}{0.0031} & 19.44 & \meanstd{0.0241}{0.0059} & 19.86 & $38.3\times$ & \meanstd{0.0225}{0.0082} & 20.76 & $1.1\times$ & \meanstd{0.0220}{0.0069} & 19.86 & $1.1\times$ \\
    11 & 2 &  & 156 & \meanstd{44.43}{0.7216} & 325.50 & \meanstd{0.3617}{0.0828} & 348.49 & $122.8\times$ & \meanstd{0.3285}{0.0769} & 369.19 & $1.1\times$ & \meanstd{0.3202}{0.0831} & 355.48 & $1.1\times$ \\
    & & & & &
    & \multicolumn{2}{r}{\textbf{Geomean}} & {$\mathbf{66.7}\times$} &  \multicolumn{2}{r}{\textbf{Geomean}} & {$\mathbf{1.1}\times$}& \multicolumn{2}{r}{\textbf{Geomean}}  & {$\mathbf{1.1}\times$} \\
    \hline
    \multicolumn{2}{c}{Module} & \multicolumn{1}{c}{$n_i$} & \multicolumn{10}{c}{{Collections-C Benchmark}} \\
    \hline
    \multicolumn{2}{c}{Array} & 21 & 35 & \meanstd{27.01}{0.1914} & 32.90 & \meanstd{7.18}{0.0633} & 9.21 & $3.8\times$ & \meanstd{7.21}{0.0448} & 9.23 & $1.0\times$ & \meanstd{7.23}{0.0774} & 9.30 & $1.0\times$ \\
    \multicolumn{2}{c}{List} & 32 & 121 & \meanstd{15.00}{0.1431} & 18.25 & \meanstd{3.65}{0.0222} & 7.92 & $4.1\times$ & \meanstd{3.59}{0.0218} & 7.92 & $1.0\times$ & \meanstd{3.65}{0.0271} & 7.95 & $1.0\times$ \\
    \multicolumn{2}{c}{Slist} & 33 & 33 & \meanstd{13.64}{0.2055} & 14.20 & \meanstd{3.34}{0.0382} & 4.39 & $4.1\times$ & \meanstd{3.45}{0.0393} & 4.51 & $1.0\times$ & \meanstd{3.42}{0.0408} & 4.46 & $1.0\times$ \\
    \multicolumn{2}{c}{RingBuffer} & 3 & 3 & \meanstd{3.78}{0.0866} & 3.91 & \meanstd{0.9563}{0.0138} & 1.08 & $4.0\times$ & \meanstd{0.9240}{0.0085} & 1.04 & $1.0\times$ & \meanstd{1.01}{0.0254} & 1.13 & $0.9\times$ \\
    \multicolumn{2}{c}{Queue} & 4 & 19 & \meanstd{75.87}{1.11} & 76.02 & \meanstd{18.09}{0.6380} & 20.50 & $4.2\times$ & \meanstd{19.88}{1.14} & 22.39 & $0.9\times$ & \meanstd{17.92}{0.2644} & 20.27 & $1.0\times$ \\
    \multicolumn{2}{c}{Treeset} & 6 & 31 & \meanstd{97.62}{1.29} & 99.13 & \meanstd{27.88}{0.3734} & 33.53 & $3.5\times$ & \meanstd{28.24}{0.5709} & 33.81 & $1.0\times$ & \meanstd{27.79}{0.2857} & 33.22 & $1.0\times$ \\
    \multicolumn{2}{c}{Treetable} & 13 & 55 & \meanstd{124.51}{2.51} & 126.51 & \meanstd{38.14}{0.2135} & 45.34 & $3.3\times$ & \meanstd{38.21}{0.1065} & 45.54 & $1.0\times$ & \meanstd{37.60}{0.2809} & 44.74 & $1.0\times$ \\
    \multicolumn{2}{c}{Deque} & 31 & 46 & \meanstd{104.38}{1.55} & 122.02 & \meanstd{24.24}{0.0522} & 28.11 & $4.3\times$ & \meanstd{24.26}{0.0543} & 28.11 & $1.0\times$ & \meanstd{25.07}{0.0482} & 28.88 & $1.0\times$ \\
    & & & & &
    & \multicolumn{2}{r}{\textbf{Geomean}} & {$\mathbf{3.9}\times$}&  \multicolumn{2}{r}{\textbf{Geomean}} & {$\mathbf{1.0}\times$} & \multicolumn{2}{r}{\textbf{Geomean}}  & {$\mathbf{1.0}\times$} \\
    \hline
    \multicolumn{3}{c}{$n_{\mathit{node}}$} & \multicolumn{10}{c}{{Arithmetic Evaluator}} \\
    \hline
     & 2000 &  & 8 & \meanstd{4.15}{0.3104} & 4.39 & \meanstd{0.3251}{0.0074} & 0.7376 & $12.8\times$ & \meanstd{0.2222}{0.0030} & 0.6313 & $1.5\times$ & \meanstd{0.1181}{0.0036} & 0.5260 & $2.8\times$ \\
     & 2000 &  & 16 & \meanstd{7.68}{0.0280} & 8.54 & \meanstd{0.6474}{0.0075} & 2.34 & $11.9\times$ & \meanstd{0.2391}{0.0022} & 1.92 & $2.7\times$ & \meanstd{0.1278}{0.0009} & 1.80 & $5.1\times$ \\
     & 2000 &  & 32 & \meanstd{16.51}{0.6243} & 20.11 & \meanstd{1.29}{0.0079} & 8.36 & $12.8\times$ & \meanstd{0.2674}{0.0026} & 7.34 & $4.8\times$ & \meanstd{0.1535}{0.0020} & 7.19 & $8.4\times$ \\
     & 4000 &  & 8 & \meanstd{10.12}{0.8170} & 10.41 & \meanstd{0.7055}{0.0127} & 1.46 & $14.3\times$ & \meanstd{0.4668}{0.0060} & 1.21 & $1.5\times$ & \meanstd{0.2618}{0.0049} & 1.01 & $2.7\times$ \\
     & 4000 &  & 16 & \meanstd{19.58}{0.6480} & 20.72 & \meanstd{1.39}{0.0109} & 4.50 & $14.1\times$ & \meanstd{0.4953}{0.0045} & 3.58 & $2.8\times$ & \meanstd{0.2787}{0.0052} & 3.31 & $5.0\times$ \\
     & 4000 &  & 32 & \meanstd{44.31}{3.13} & 49.23 & \meanstd{2.76}{0.0275} & 15.52 & $16.1\times$ & \meanstd{0.5433}{0.0048} & 13.47 & $5.1\times$ & \meanstd{0.3252}{0.0058} & 13.14 & $8.5\times$ \\
     & 8000 &  & 8 & \meanstd{29.46}{0.7530} & 29.85 & \meanstd{1.07}{0.0230} & 1.98 & $27.5\times$ & \meanstd{0.6163}{0.0044} & 1.53 & $1.7\times$ & \meanstd{0.4033}{0.0038} & 1.30 & $2.7\times$ \\
     & 8000 &  & 16 & \meanstd{65.37}{2.57} & 67.03 & \meanstd{3.16}{0.0580} & 9.18 & $20.7\times$ & \meanstd{1.05}{0.0116} & 7.08 & $3.0\times$ & \meanstd{0.6012}{0.0047} & 6.33 & $5.3\times$ \\
     & 8000 &  & 32 & \meanstd{145.49}{9.18} & 152.22 & \meanstd{5.82}{0.0148} & 29.51 & $25.0\times$ & \meanstd{1.15}{0.0082} & 25.03 & $5.1\times$ & \meanstd{0.7410}{0.0105} & 24.22 & $7.9\times$ \\
    & & & & &
    & \multicolumn{2}{r}{\textbf{Geomean}} & {$\mathbf{16.5}\times$} &  \multicolumn{2}{r}{\textbf{Geomean}} & {$\mathbf{2.8}\times$} & \multicolumn{2}{r}{\textbf{Geomean}}  & {$\mathbf{4.8}\times$} \\
    \hline
    $n_{\mathit{comp}}$ & $n_{\mathit{sym}}$ & \multicolumn{1}{c}{$n_{\mathit{size}}$} & \multicolumn{10}{c}{{Quicksort}} \\
    \hline
    16 & 2 & 20 & 315 & \meanstd{3.72}{0.0736} & 18.32 & \meanstd{0.3131}{0.0012} & 9.80 & $11.9\times$ & \meanstd{0.2472}{0.0016} & 9.74 & $1.3\times$ & \meanstd{0.2448}{0.0023} & 9.88 & $1.3\times$ \\
    16 & 2 & 30 & 767 & \meanstd{16.69}{0.8284} & 73.97 & \meanstd{1.39}{0.0038} & 35.00 & $12.0\times$ & \meanstd{1.02}{0.0036} & 34.94 & $1.4\times$ & \meanstd{1.03}{0.0047} & 35.03 & $1.3\times$ \\
    16 & 2 & 40 & 1415 & \meanstd{51.44}{0.5677} & 208.43 & \meanstd{4.17}{0.0533} & 88.36 & $12.3\times$ & \meanstd{3.00}{0.0151} & 86.46 & $1.4\times$ & \meanstd{3.03}{0.0429} & 86.77 & $1.4\times$ \\
    16 & 2 & 50 & 2270 & \meanstd{131.72}{2.03} & 483.66 & \meanstd{9.90}{0.0606} & 177.44 & $13.3\times$ & \meanstd{7.29}{0.0231} & 176.36 & $1.4\times$ & \meanstd{7.40}{0.0225} & 176.48 & $1.3\times$ \\
    25724 & 3 & 5 & 112 & \meanstd{1085.52}{20.80} & 1088.07 & \meanstd{0.4650}{0.2527} & 1.73 & $2334.5\times$ & \meanstd{0.5711}{0.3127} & 2.06 & $0.8\times$ & \meanstd{0.1072}{0.0532} & 1.37 & $4.3\times$ \\
    25724 & 3 & 10 & 862 & - & \TO & \meanstd{4.34}{0.0556} & 21.38 & - & \meanstd{9.79}{0.0333} & 27.98 & $0.4\times$ & \meanstd{1.90}{0.0570} & 19.24 & $2.3\times$ \\
    & & & & &
    & \multicolumn{2}{r}{\textbf{Geomean}} & {$\mathbf{35.3}\times$}&  \multicolumn{2}{r}{\textbf{Geomean}} & {$\mathbf{1.0}\times$} & \multicolumn{2}{r}{\textbf{Geomean}}  & {$\mathbf{1.8}\times$} \\
    \Xhline{1.2pt}
    \multicolumn{8}{r}{\textbf{Overall Geomean}}
    & \multicolumn{1}{c}{$\mathbf{29.4}\times$}&  \multicolumn{2}{r}{\textbf{Overall Geomean}} & \multicolumn{1}{c}{$\mathbf{1.3}\times$} & \multicolumn{2}{r}{\textbf{Overall Geomean}}  &  $\mathbf{1.5}\times$ \\
  \end{tabular}%
  }
  \flushleft
  {
  \small
  Note: $T_{\mathit{exec}}$ reports the mean instruction-execution time over 5 runs;
  $\sigma_{\mathit{exec}}$ reports their standard deviations.
  Timings are measured in seconds.
  Time out (TO) is set to 2 hours for end-to-end execution. \\
  RQ1 speedup $S_{\mathit{staging}} =
  T_{\mathit{exec}}^{\WASP} / T_{\mathit{exec}}^{\mathit{noreuse}}$ compares
  \tool without snapshot-reuse with \WASP;\\
  RQ2 speedup $S_{\mathit{snapshot}} =
  T_{\mathit{exec}}^{\mathit{noreuse}} /
  T_{\mathit{exec}}^{\mathit{snapshot}}$ compares \emph{uniform} snapshot-reuse
  with no-reuse;\\
  RQ2 speedup $S_{\mathit{heuristic}} = T_{\mathit{exec}}^{\mathit{noreuse}} /
  T_{\mathit{exec}}^{\mathit{heuristic}}$ compares \emph{heuristic} snapshot-reuse
  with no-reuse.
  }

  \vspace{-1.5em}
\end{table*}

\begin{figure}
  \centering
  \begin{subfigure}[t]{0.49\columnwidth}
    \vspace{0pt}
    \centering
    \includegraphics[width=\linewidth]{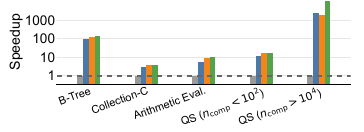}
    \vspace{-1.5em}
    \caption{Per-benchmark speedup. Bars show \WASP (1$\times$) and three \tool
    configurations: no snapshot reuse, uniform snapshot reuse, and heuristic
    snapshot reuse. Higher is better; QS denotes quicksort.}
    \label{fig:speedup-by-benchmark}
  \end{subfigure}\hfill
  \begin{subfigure}[t]{0.49\columnwidth}
    \vspace{0pt}
    \centering
    \includegraphics[width=\linewidth]{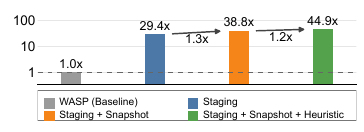}
    \vspace{-1.5em}
    \caption{Overall average speedup, computed as the geometric mean of all the benchmarks
      speedups across 4 benchmark groups. Arrows show the speedup factor from each
      \tool configuration to the next.}
    \label{fig:speedup-mean}
  \end{subfigure}
  \vspace{-0.5em}
  \caption{Speedup from staging, snapshot reuse and snapshot reuse heuristics. }
  \label{fig:speedup}
  \vspace{-1em}
\end{figure}

\subsection{RQ1: \tool Performance Compared to Interpretation-Based Approaches}

\subsubsection*{Method}
To isolate the benefits of staging from snapshot-reuse, we disable
snapshot-reuse in \tool for RQ1, forcing each execution to restart from the
program entry, as in \WASP. We report the mean instruction-execution time and
end-to-end execution time, with a two-hour wall-clock timeout per run. For
\WASP, the measurements exclude WebAssembly module parsing.

We configure the B-Tree benchmark using two parameters: the numbers of ordered
and unordered symbolic inputs, denoted by $n_o$ and $n_u$, respectively.
The extended Collections-C benchmark contains $n_i$ test cases for each data structure,
listed in \Cref{tab:rq1-uniform}. We compile all test cases to WebAssembly and
perform concolic execution on the resulting modules. For every RQ1 benchmark,
we ensure that \tool and \WASP explore the same number of paths.

The two benchmark suites play complementary roles in assessing performance
across synthetic and realistic workloads. 
The extended B-Tree benchmark provides a controlled setting for systematically
varying the state-space by imposing different ordering constraints on symbolic
inputs.
For example, $a > b$ excludes inputs where $a\le b$, reducing the feasible input space.
Collections-C provides more realistic workloads: its Wasm programs are compiled
from a C implementation of various collection data structures that are widely
used in real-world applications.

\subsubsection*{Results}

\Cref{tab:rq1-uniform} shows the results.
We compute the speedup of staging $S_{\mathit{staging}}$ as
$T_{\mathit{exec}}^{\mathit{\WASP}} / T_{\mathit{exec}}^{\mathit{noreuse}}$.
We observe that \tool consistently and substantially outperforms \WASP in
instruction-execution time across all configurations of the extended B-Tree
and Collections-C benchmarks (\Cref{tab:rq1-uniform}).
On average, \tool achieves a $66.7\times$ speedup over \WASP on the B-Tree
benchmark, with the highest $652.5\times$ speedup excluding the timed-out
configuration $(n_o, n_u) = (10, 3)$.
On the Collections-C benchmark, \tool achieves a $3.9\times$ speedup.
Overall, across 169 test programs spanning two benchmark groups,
\tool achieves an average speedup of $33.4\times$.
On the arithmetic-evaluator benchmark and the quicksort benchmark,
\tool achieves an average speedup of $21.6\times$.

On the B-Tree benchmark, \WASP exhibits increasingly severe slowdowns as
execution time grows. We hypothesize that this degradation stems largely from
its small-step interpretation and garbage collection (GC) in OCaml. The
performance gains of
\tool come primarily from three sources:
(1) Compilation eliminates the interpretation overhead of repeatedly
dispatching instructions.
(2) By leveraging the Wasm CPS semantics \cite{DBLP:conf/sfp/WeiBZZ25}, the
generated C++ code translates control-flow instructions directly into function
calls, whereas \WASP evaluates control flow through administrative
instructions.
(3) And lastly, the generated code avoids GC and uses efficient data structures to
represent execution state, including the stack and environment.

\subsubsection*{Summary}
\Cref{fig:speedup-by-benchmark} summarizes the speedup results for each
benchmark group.
Overall, the geometric mean of \tool's speedups over \WASP across all four
benchmark groups is $29.4\times$.

\subsection{RQ2: When Are Snapshot Reuse and Our Heuristic Effective?}
\label{sec:rq2}

\subsubsection*{Method}
To evaluate snapshot reuse and our heuristic, we compare configurations of
\tool with these features enabled or disabled. In general, the net benefit of snapshot
reuse depends on two factors: the execution it avoids and the cost of
instantiating snapshots. Although the RQ1 benchmarks support a broad performance
evaluation, they do not isolate this trade-off. We therefore use two controlled
benchmarks (an arithmetic evaluator and quicksort) to vary these factors
independently. For reference, we also report \WASP's performance.

The arithmetic-evaluator benchmark is parameterized by the input-expression AST
size ($n_{\mathit{node}}$ in \Cref{tab:rq1-uniform}) and the number of explored
paths. Increasing the expression size increases the execution depth of each path
and thus the number of executed instructions.  Because path forking occurs late
in execution in this benchmark, increasing the number of paths would increase
sharing in the execution tree, allowing snapshot reuse to skip more common
execution.

Quicksort may reorder array elements but does not construct new concrete or
symbolic values. We exploit this property to control symbolic-state complexity
through the AST sizes of the symbolic inputs. Each symbolic input can be initialized
as either an atomic symbolic variable or a complex symbolic arithmetic
expression. Larger expressions make snapshots more expensive to copy and
instantiate. \Cref{tab:rq1-uniform} shows the benchmark parameters: the
syntactic complexity of symbolic inputs $n_{\mathit{comp}}$, the input
array size $n_{\mathit{size}}$, and the number of symbolic elements $n_{\mathit{sym}}$.
The remaining $n_{\mathit{size}} - n_{\mathit{sym}}$
elements are initialized with distinct concrete values.
We define $n_{\mathit{comp}}$ as the sum of the syntactic sizes of all symbolic
values in the array.
For example, storing a 32-bit symbolic value $x$ as 4 bytes in the Wasm
memory requires 4 SMT $\mathsf{extract}_{hi,lo}(x)$ expressions which have size
2 each, so an atomic symbolic variable has size 8.
We can deliberately construct symbolic states of arbitrary complexity by
repeatedly applying arithmetic operations to the initial symbolic values.

\subsubsection*{Results}
\Cref{tab:rq1-uniform} shows the results.
The columns $T_{\mathit{exec}}^{\mathit{noreuse}}$,
$T_{\mathit{exec}}^{\mathit{snapshot}}$, and $T_{\mathit{exec}}^{\mathit{heuristic}}$
report instruction-execution times without snapshot reuse, with uniform snapshot
reuse, and with heuristic snapshot reuse, respectively.
We report three speedups to quantify the benefits of avoiding re-execution:
$S_{\mathit{staging}}$, the speedup of staging over the \WASP baseline;
$S_{\mathit{snapshot}}$, the speedup of \emph{uniform} snapshot reuse over no-reuse;
and $S_{\mathit{heuristic}}$, the speedup of \emph{heuristic} snapshot reuse over
no-reuse.

On the B-Tree and Collections-C benchmarks, both uniform and heuristic snapshot
reuse yield modest average speedups of $1.1\times$, compared to \tool with
snapshot reuse disabled.
Some configurations instead exhibit slight slowdowns, with
$S_{\mathit{snapshot}}$ as low as $0.9\times$ on the Queue benchmark, and
$S_{\mathit{heuristic}}$ as low as $0.9\times$ on the RingBuffer benchmark.
Although snapshot reuse avoids re-executing shared prefixes, it does not always
reduce the total execution time because the cost of creating and resuming
snapshots is comparable to that of re-executing those prefixes.

On the arithmetic-evaluator benchmark, \emph{heuristic} snapshot reuse achieves an
average $4.8\times$ speedup over no-reuse.
It also outperforms \emph{uniform} snapshot reuse by an average of $1.7\times$.
Although snapshot incur overhead, its better reuse decisions
consistently reduce overall execution time.

The quicksort results exhibit a different pattern. With atomic symbolic
expressions as inputs (\eg, $n_{\mathit{comp}} \leq 10^2$),
uniform snapshot reuse remains beneficial because snapshots can bypass shared
execution prefixes at negligible resumption cost.
With more complex symbolic states (\eg, $n_{\mathit{comp}} > 10^4$), however,
uniform reuse slows execution relative to no-reuse, yielding an average
$S_{\mathit{snapshot}}$ of $0.6\times$. For short executions, the cost of
copying and resuming snapshots containing large symbolic expressions outweighs
the cost of re-executing the shared prefix.
However, the heuristic effectively avoids these unfavorable cases, reusing
snapshots only when the saved execution time justifies the overhead and yielding
a $3.1\times$ speedup over no-reuse.

\subsubsection*{Summary}
Uniform snapshot creation and restoration can outweigh the saved execution time,
causing slowdowns in some cases.  Our heuristic identifies these cases and makes
more effective reuse decisions, consistently outperforming uniform snapshot
reuse.
\Cref{fig:speedup-mean} shows the average speedups from snapshot reuse.
Beyond staging's $29.4\times$ speedup over \WASP, uniform snapshot reuse adds
$1.3\times$ speedup (\ie $38.8\times$ vs \WASP),
and heuristics adds another $1.2\times$ speedup over uniform snapshot reuse
(\ie, $44.9\times$ vs \WASP).

\subsection{RQ3: Is \tool effective in finding real bugs?}

\subsubsection*{Method}
To evaluate the effectiveness of \tool in detecting real bugs in practical
programs, we conduct experiments on two \emph{buggy programs} from Collections-C.
These programs were previously used to evaluate the bug-finding effectiveness of
\WASP \cite{DBLP:conf/ecoop/MarquesS0A22} and Gillian \cite{DBLP:conf/pldi/SantosMAG20}.
Both contain memory-safety violations that may cause memory corruption and
thus pose potential security risks.
For our evaluation, we compile the Collections-C test cases from C to
WebAssembly and run concolic execution on the resulting Wasm modules.
We also run \WASP as a reference baseline for comparison.

\begin{table}[H]\small
  \centering
  \vspace{-0.5em}
  \caption{Bug detection results on the \emph{buggy version} of Collections-C. Timings are measured in seconds.}
  \label{tab:rq4-bugs}
  \vspace{-0.5em}
  \resizebox{0.8\columnwidth}{!}{  \begin{tabular}{c c c c c c c c c}
    \toprule
    Test &
    $n_{\mathit{paths}}^{\text{\tool}}$ &
    $n_{\mathit{paths}}^{\text{\WASP}}$ &
    Finds bug &
    $T_{\mathit{exec}}^{\text{\tool}}$ &
    $T_{\mathit{total}}^{\text{\tool}}$ &
    $T_{\mathit{exec}}^{\text{\WASP}}$ &
    $T_{\mathit{total}}^{\text{\WASP}}$ \\
    \midrule
    \texttt{array\_test\_remove} & 13 & 1 & Yes  & 0.0153 & 0.2663 & 0.0042 & 0.0155 \\
    \texttt{list\_test\_zipIterAdd} & 2028 & 1  & Yes & 3.71 & 12.2 & 0.0058 & 0.0058 \\
    \bottomrule
  \end{tabular}  }
  \vspace{-0.5em}
\end{table}

\subsubsection*{Results}

\Cref{tab:rq4-bugs} shows that both \tool and \WASP detect the bugs.
\tool explores more paths and takes longer than \WASP because they use different
concolic execution strategies: \WASP stops upon detecting a bug, whereas \tool
continues exploring more paths for higher code coverage.
These results show that \tool can effectively find real bugs in practical
programs.

\subsection{Threats to Validity}

\subsubsection*{Generalizability of Snapshot-Reuse.}
In RQ2, we evaluated snapshot reuse and our heuristic using relatively small
benchmarks with short execution times.
These programs also do not contain external calls.
In the \Cref{sec:heuristic}, we use the size of symbolic expressions to estimate
the instantiation time of snapshot; it is an approximation of the actual
instantiation time, which may be affected by other factors such as
the structure of symbolic expressions and the underlying theory being used.

\subsubsection*{Unsupported Language Features.}
As a prototype, \tool supports a substantial subset of Wasm, but not its full
feature set. Applying \tool to real-world programs may therefore require users
to implement unsupported features, such as external calls and SIMD
instructions.

\subsubsection*{Compilation Time}
In the evaluation, we compared only the Wasm instruction-execution time for both
\tool and \WASP. 
However, when a program's execution time is short, \tool's compilation time may
be comparable to or even longer than its execution time.
When using \tool in practice, one should weigh the compilation cost against the
expected savings in execution time to ensure a positive overall benefit.

\section{Discussion} \label{sec:discussion}

\subsubsection*{Instrumentation vs. Staging}
Instrumentation is another popular approach to implementing efficient concolic
execution engines \cite{DBLP:conf/uss/Yun0XJK18, poeplau2020symcc,
EURECOM+6459}. It is a source-to-source transformation that generates an
instrumented program for performing concolic execution. However, instrumentation
and staging differ in several respects when used to implement concolic execution
engines.

To achieve efficiency without interpretation overhead, instrumentation requires
the underlying target language to already have an efficient compiler-based
implementation, such as C or LLVM IR. Existing commodity compiler
infrastructure can then be leveraged to implement the instrumentation and
generate instrumented code. In contrast, our approach is semantics-first: it
starts from a concrete semantics and derives a concolic interpreter for the
target language. The efficiency gain comes from eliminating interpretation
overhead through staging. In this sense, we use staging to implement a compiler
for a non-standard execution semantics.

One consequence of this difference is that instrumentation-based concolic
execution preserves the original program's concrete execution, including its
control flow and native state representation, whereas staging-based concolic
execution must define these aspects explicitly in the semantics.
The benefit is that staging-based concolic execution offers greater flexibility
in customizing and manipulating the program's control and state, which is useful
for implementing features such as snapshot reuse. Our approach takes advantage
of this flexibility by representing the program's control as continuations at
the meta-level, thereby enabling flexible snapshot reuse.
In contrast, the reliance on the underlying language's concrete execution makes
snapshot reuse challenging to implement by instrumentation, because it requires
reifying the program's control and concrete states as manipulable data
structures, which is not always straightforward or even possible.

In general, instrumentation can also be understood as partial evaluation
\cite{DBLP:conf/rv/FeiginM10}. \citet{DBLP:conf/rv/FeiginM10}
characterize the efficiency of instrumentation by drawing on ideas from Jones
Optimality~\cite{DBLP:books/daglib/0072559}.

\subsubsection*{Adopt Our Approach in Other Languages/IRs}

We instantiate our proposed architecture for Wasm, but the underlying methodology
is language-independent.
To adopt it for another language, the implementer can start from a
concrete definitional interpreter in CPS for that language.
Following the mechanical steps in \Cref{sec:concolic,sec:ce-compiler}, the
implementer can extend the concrete domains with their symbolic counterparts to
obtain a concolic interpreter and then refactor it into a staged one.

Additionally, adopting the approach would entail building the necessary runtime
support for state representation, symbolic operations, SMT solver interaction,
and path scheduling.
The generated code from specializing the concolic interpreter should work with
the runtime support to perform concolic execution of the input program.

\section{Related Work} \label{sec:related}

\subsubsection*{\textbf{Concolic Execution}}
Concolic execution \cite{10.1145/1065010.1065036, DBLP:conf/sigsoft/SenMA05} is a
white-box testing technique, which guides symbolic execution by concrete execution.
\citet{DBLP:conf/esop/YouFD21} extend the idea of concolic execution
to higher-order programs where functions are first-class values.
Hybrid concolic execution \cite{DBLP:conf/icse/MajumdarS07} combines fuzzing
and concolic testing, which has been resulted in various tools
\cite{DBLP:conf/uss/Yun0XJK18, DBLP:conf/ndss/StephensGSDWCSK16}.
Often conflated in the literature, KLEE \cite{10.5555/1855741.1855756} performs
fork-based symbolic execution, in contrast to concolic execution tools such as
CUTE \cite{DBLP:conf/sigsoft/SenMA05}, which would require concrete input seeds.

In the Wasm ecosystem, \Manticore \cite{DBLP:conf/kbse/MossbergMHGGFBD19}
and \SeeWasm \cite{DBLP:conf/issta/HeZGWP0WC024} are both Python-based symbolic
execution engines, while \Owi
\cite{DBLP:journals/corr/abs-2412-06391} supports parallel symbolic execution of
Wasm programs. None of these tools supports concolic execution.
To our knowledge, \WASP \citet{DBLP:conf/ecoop/MarquesS0A22} was the only concolic
execution engine for Wasm, implemented by extending the official Wasm
interpreter.
All of these approaches rely on interpretation and therefore incur substantial
interpretation overhead.

\subsubsection*{\textbf{Reuse in Concolic Execution}}
\QSYM \cite{DBLP:conf/uss/Yun0XJK18} performs concolic execution via dynamic binary
instrumentation, thus avoids the interpretation overhead over IRs compared to
traditional interpretation-based engines.
The authors of \QSYM argue that its performant execution makes it
unnecessary to use an inefficient snapshot mechanism to avoid re-execution,
especially compared to previous tools such as Driller \cite{DBLP:conf/ndss/StephensGSDWCSK16}.
However, it is worth noting to explain the causes of Driller's unsatisfying
performance.
Driller uses angr \cite{DBLP:conf/ndss/Shoshitaishvili15} as its underlying
concolic execution engine. Angr first translates binary code into Valgrind's
VEX IR \cite{DBLP:conf/pldi/NethercoteS07}, which will be \emph{interpreted}
to perform concolic execution.
The indirection layer of interpretation makes it easy to implement
``snapshots'', but also contributes to the slowdown.
This paper attempts to bring the best of two worlds, \ie, fast execution
via compilation and the ability to reuse symbolic states via continuations.

In hybrid concolic execution, a fuzzer produces new inputs via mutation.
However, the correspondence between those inputs and the negation point is
unknown to the concolic execution engine, thus preventing effective reuse.

\subsubsection*{\textbf{Code Generation for Symbolic/Concolic Execution}}
Our approach achieves compilation via specializing a definitional concolic interpreter.
A commonly used implementation strategy for concolic execution is
referred as instrumentation, for example, \QSYM \cite{DBLP:conf/uss/Yun0XJK18} uses Intel's
Pin to perform instruction-level dynamic binary instrumentation, \SymCC
\cite{poeplau2020symcc} lifts the level of instrumentation to LLVM's IR, and
\textsc{SymQEMU} \cite{EURECOM+6459} instruments over QEMU's IR.
Compared to our approach, all these instrumentation-based concolic engines lack
the ability to control the execution and require re-execution.
\GenSym \cite{icse23,10.1145/3428232} uses staging to achieve efficient symbolic execution for
LLVM IR and inspires our work, but does not support concolic execution.

\section{Conclusion} \label{sec:conclusion}

In this paper, we presented a novel approach to efficient concolic execution by
compiling a definitional concolic interpreter written in CPS
using multi-stage programming.  Our approach offers three key benefits.
First, by expressing concolic execution as a definitional interpreter, we
reduce the implementation effort compared to building a compiler from
scratch. Second, leveraging multi-stage programming eliminates interpretation
overhead, yielding highly efficient generated code. Third, representing
control flow with continuations allows us to capture snapshots and avoid costly
re-execution when exploring new paths.
We further introduce a heuristic for deciding when to reuse continuations, which
improves performance in practice.  We implemented this approach in a prototype
concolic-execution compiler \tool for WebAssembly programs. Our evaluation
shows that \tool significantly outperforms prior systems.

\section*{Data-Availability Statement}
The paper's artifact is available at \url{https://zenodo.org/records/21769340}
\cite{genwasym_artifact}, which contains the implementation of \tool, the
benchmark suite used in the paper, and the evaluation scripts.
The \tool is publicly maintained at
\url{https://github.com/Generative-Program-Analysis/GenSym/}.

\begin{acks}
We thank the reviewers of OOPSLA 2026 and PLDI 2026 for their meticulous reviews
and constructive suggestions.
We thank Léo Andrès for helpful discussions about Owi and WASP.
We also would like to thank the members of the Tufts Programming Languages
Group (TuPL) and the audience of the WebAssembly Workshop at
OOPSLA/ICFP 2025 for their valuable feedback on this work.
\end{acks}

\bibliographystyle{ACM-Reference-Format}
\bibliography{references}

\clearpage
\appendix
\renewcommand{\theHequation}{appendix.\Alph{section}.\arabic{equation}}
\section{Complete Semantics and Algorithms}
\label{sec:appendix:full-definitions}

This appendix contains the complete versions of the semantics and algorithms
excerpted or omitted in the paper:
\begin{itemize}
  \item \Cref{fig:concolic:semantics:full} gives the full concolic continuation
        semantics corresponding to the excerpt in \Cref{fig:concolic:semantics}.
  \item \Cref{fig:concolic:driver:full} gives the complete concolic scheduler,
        including the definition of \textsc{Negate}, corresponding to
        \Cref{fig:concolic:driver}.
  \item \Cref{fig:staging:semantics:full} gives the full staged concolic
        continuation semantics corresponding to the excerpt in
        \Cref{fig:staging:semantics}.
  \item \Cref{fig:staging:driver} gives the staged concolic scheduler used in
        \Cref{sec:compilation:scheduler}.
  \item \Cref{fig:snapshot:driver:full} gives the complete concolic snapshot
         scheduler corresponding to \Cref{fig:snapshot:driver}.
  \item \Cref{fig:heuristic:driver} gives the concolic snapshot scheduler
        adapted to use the heuristic described in \Cref{sec:heuristic}.
\end{itemize}

\clearpage
\begin{figure}[p]\footnotesize
\judgement{Semantic Domains}{}
\begin{equation}
  \nonumber
  \begin{alignedat}{3}
    op^\#_2  & \in \{ +^\#, -^\#, \ldots \} && \hspace{2em} op^\#1 \in \{ -^\#, \dots \} \\
    x         & \in \HL{\Typ{SymVar}}        && := \Typ{Identifier} \\
    s         & \in \HL{\Typ{Symbolic}} && := v \mid \Typ{Sym}(x) \mid s_1 ~ op^\#_2 ~  s_2 \mid op^\#_1 ~ s_1\\
    \pi^\#    & \in \HL{\Typ{PC^\#}}       && = \Typ{List[Symbolic]} \\
  \end{alignedat}
  \qquad
  \begin{alignedat}{3}
    \sigma^\# & \in \Typ{Stack}^\# && = \Typ{List[Value \times \HL{\Typ{Symbolic}}]} \\
    \rho^\#   & \in \Typ{Env}^\# && = \Typ{List[Value \times \HL{\Typ{Symbolic}}]} \\
    \kappa^\# & \in \Typ{Cont^\#} && = (\Typ{Stack}^\# \times \Typ{Env}^\# \times \HL{\Typ{PC^\#}}) \to \Ans \\
    \theta^\# & \in \Typ{Trail^\#} && = \Typ{List[Cont^\#]}
  \end{alignedat}
\end{equation}
\judgement{Concolic Continuation Semantics}{}
  \begin{alignat*}{3}
    & \Concdeno{\cdot} : \mathrlap{\Typ{List[Inst]} \rightarrow (\Typ{Stack^\#} \times \Typ{Env^\#} \times \HL{\Typ{PC^\#}} \times \Typ{Trail^\#} \times \Typ{Cont^\#}) \rightarrow \Typ{Ans}} \\
    & \Concdeno{\mathit{nil}} (\sigma^\#, \rho^\#, \HL{\pi^\#}, \kappa^\#, \theta^\#) && = \kappa^\#(\sigma^\#, \rho^\#, \HL{\pi^\#}) \\
    & \Concdeno{\Inst{nop} :: \mathit{rest}} (\sigma^\#, \rho^\#, \HL{\pi^\#}, \kappa^\#, \theta^\#) && =
      \Concdeno{\mathit{rest}} (\sigma^\#, \rho^\#, \HL{\pi^\#}, \kappa^\#, \theta^\#) \\
    & \Concdeno{t.\Inst{const} ~ c :: \mathit{rest}} (\sigma^\#, \rho^\#, \HL{\pi^\#}, \kappa^\#, \theta^\#) && =
      \Concdeno{\mathit{rest}} (\la c, \HL{c}\ra :: \sigma^\#, \rho^\#, \HL{\pi^\#}, \kappa^\#, \theta^\#) \\
    & \Concdeno{t.\Inst{add} :: \mathit{rest}} (\la v_1, \HL{s_1}\ra :: \la v_2, \HL{s_2}\ra :: \sigma^\#, \rho^\#, \HL{\pi^\#}, \kappa^\#, \theta^\#) && =
      \Concdeno{\mathit{rest}} (\la v_1 + v_2, \HL{s_1 ~ +^\# ~ s_2}\ra :: \sigma^\#, \rho^\#, \HL{\pi^\#}, \kappa^\#, \theta^\#) \\
    & \Concdeno {\Inst{local.get} ~ x :: \mathit{rest}} (\sigma^\#, \rho^\#, \HL{\pi^\#}, \kappa^\#, \theta^\#) && =
      \Concdeno{\mathit{rest}}(\rho^\#(x) :: \sigma^\#, \rho^\#, \HL{\pi^\#}, \kappa^\#, \theta^\#) \\
    & \Concdeno{\Inst{local.set} ~ x :: \mathit{rest}}(\la v, \HL{s}\ra :: \sigma^\#, \rho^\#, \HL{\pi^\#}, \kappa^\#, \theta^\#) && =
      \Concdeno{\mathit{rest}}(\sigma^\#, \rho^\#[x \mapsto \la v, \HL{s}\ra], \HL{\pi^\#}, \kappa^\#, \theta^\#)\\
    & \Concdeno{\Inst{block} ~ (t^m \to t^n) ~ \es :: \mathit{rest}}(\sigma^\#_{\mathit{arg}} \pconcat{m} \sigma^\#, \rho^\#, \HL{\pi^\#}, \kappa^\#, \theta^\#) && = \\
       & && \magic \Typ{let} ~ \kappa^\#_1 \coloneqq \lambda (\sigma^\#_1, \rho^\#_1, \HL{\pi^\#_1}). \Concdeno{\mathit{rest}}(\truncate{\sigma^\#_1}{n} \mathbin{+\!\!+} \sigma^\#, \rho^\#_1, \HL{\pi^\#_1}, \kappa^\#, \theta^\#) ~ \Typ{in} \\
       & && \magic \Concdeno{\es}(\sigma^\#_{\mathit{arg}}, \rho^\#, \HL{\pi^\#}, \kappa^\#_1, \kappa^\#_1 :: \theta^\#) \\
    & \Concdeno{\Inst{loop} ~ (t^m \to t^n) ~ \es :: \mathit{rest}}(\sigma^\#_{\mathit{arg}} \pconcat{m} \sigma^\#, \rho^\#, \HL{\pi^\#}, \kappa^\#, \theta^\#) && = \\
       & && \magic \Typ{let} ~ \kappa^\#_1 \coloneqq \lambda (\sigma^\#_1, \rho^\#_1, \HL{\pi^\#_1}). \Concdeno{\mathit{rest}}(\truncate{\sigma^\#_1}{n} \mathbin{+\!\!+} \sigma^\#, \rho^\#_1, \HL{\pi^\#_1}, \kappa^\#, \theta^\#) ~ \Typ{in} \\
       & && \magic \Typ{fix} ~ \kappa^\#_2 \coloneqq \lambda (\sigma^\#_2, \rho^\#_2, \HL{\pi^\#_2}). \Concdeno{\es}(\truncate{\sigma^\#_2}{m}, \rho^\#_2, \HL{\pi^\#_2}, \kappa^\#_1, \kappa^\#_2 :: \theta^\#) ~ \Typ{in} \\
       & && \magic \kappa^\#_2(\sigma^\#_{\mathit{arg}}, \rho^\#, \HL{\pi^\#}) \\
    & \Concdeno{\Inst{if} ~ (t^m \to t^n) ~ \es_1 ~ \es_2 :: \mathit{rest}}(\la v, \HL{s}\ra :: \sigma^\#_{\mathit{arg}} \pconcat{m} \sigma^\#, \rho^\#, \HL{\pi^\#}, \kappa^\#, \theta^\#) && = \\
       & && \magic \Typ{let} ~ es \coloneqq \Typ{if} ~ v \equiv 0 ~ \Typ{then} ~ \es_2 ~ \Typ{else} ~ \es_1 ~ \Typ{in} \\
       & && \magic \Typ{let} ~ \HL{\pi^\#_1} \coloneqq \Typ{if} ~ v \equiv 0 ~ \Typ{then} ~ \neg s :: \pi^\# ~ \Typ{else} ~ s :: \pi^\# ~ \Typ{in} \\
       & && \magic \Typ{let} ~ \kappa^\#_1 \coloneqq \lambda (\sigma^\#_1, \rho^\#_1, \HL{\pi^\#_1}). \Concdeno{\mathit{rest}}(\truncate{\sigma^\#_1}{n} \mathbin{+\!\!+} \sigma^\#, \rho^\#_1, \HL{\pi^\#_1}, \kappa^\#, \theta^\#) ~ \Typ{in} \\
       & && \magic \Concdeno{es}(\sigma^\#_{\mathit{arg}}, \rho^\#, \HL{\pi^\#_1}, \kappa^\#_1, \kappa^\#_1 :: \theta^\#) \\
    & \Concdeno{\Inst{br} ~ \ell :: \mathit{rest}}(\sigma^\#, \rho^\#, \HL{\pi^\#}, \kappa^\#, \theta^\#) && = \theta^\#(\ell)(\sigma^\#, \rho^\#, \HL{\pi^\#}) \\
    & \Concdeno{\Inst{call} ~ x :: \mathit{rest}}(\sigma^\#_{\mathit{arg}} \pconcat{m} \sigma^\#, \HL{\pi^\#}, \rho^\#, \kappa^\#, \theta^\#) && = \\
       & && \magic \Typ{let} ~ \{ \Typ{type}: t^m \to t^n, \Typ{locals}: \mathit{ts}, \Typ{body}: \es \} \coloneqq \Typ{lookupFunc}(x) ~ \Typ{in} \\
       & && \magic \Typ{let} ~ \rho^\#_1 \coloneqq \mathsf{buildEnv}(\sigma^\#_{\mathit{arg}}, \mathit{ts}) \\
       & && \magic \Typ{let} ~ \kappa^\#_1 \coloneqq \lambda (\sigma^\#_1, \rho^\#_1, \HL{\pi^\#_1}). \Concdeno{\mathit{rest}}(\truncate{\sigma^\#_1}{n} \concat \sigma^\#, \rho^\#, \HL{\pi^\#_1}, \kappa^\#, \theta^\#) ~ \Typ{in} \\
       & && \magic \Concdeno{\es}([], \rho^\#_1, \kappa^\#_1, [\kappa^\#_1]) \\
    & \Concdeno{\Inst{return} :: \mathit{rest}}(\sigma^\#, \rho^\#, \HL{\pi^\#}, \kappa^\#, \theta^\#) && = \theta^\#.\mathsf{last}(\sigma^\#, \rho^\#, \HL{\pi^\#})
  \end{alignat*}
  \caption{The concolic semantics of \lang (complete version of
    \Cref{fig:concolic:semantics}).}\label{fig:concolic:semantics:full}
\end{figure}

\begin{algorithm}[t]
  \small
  \caption{Concolic Scheduler (complete version of
    \Cref{fig:concolic:driver})}
  \label{fig:concolic:driver:full}
  \begin{algorithmic}[1]
    \Function{\textsc{Scheduler}}{$\mathit{mod} \in \Typ{Module}$, $f \in \Typ{Identifier}$, $\mathit{input} \in \Typ{List[Symbolic]}$}
      \State $WL \gets \{\, [] \, \}$ \Comment{$WL \in \Pow{\Typ{PC^\#}}$}
      \State $Seen \gets \varnothing$ \Comment{$Seen \in \Pow{\Typ{PC^\#}}$}
      \While{$WL \neq \varnothing$}
        \State pop $\pi$ from $WL$
        \IIf{$\pi \in Seen$} \textbf{continue}
        \State $m \gets \mathit{solve}(\pi)$ \Comment{invoke SMT solver}
        \IIf{$m = \UNSAT$} \textbf{continue} \Comment{unreachable path condition}
        \State $\pi_1 \gets \Concdeno{\Inst{call} ~ f}(\mathit{instantiate}(\mathit{input}, m), [], [\kappa^\#_{halt}], \kappa^\#_{halt})$ \Comment{run concolic semantics}
        \State $Seen \gets Seen \cup \{\pi_1\}$
        \ForAll{$\pi' \in \textsc{Negate}(\pi_1)$}
          \State $WL \gets WL \cup \{\pi'\}$
        \EndFor
      \EndWhile
    \EndFunction
    \\
    \Function{\textsc{Negate}}{$\pi \in \Typ{PC^\#}$} $\rightarrow$ $\Typ{List[PC^\#]}$ \dots
      \State \textbf{match} $\pi$ \textbf{with}
      \State \quad \textbf{case} $[]$ $\rightarrow$ $[]$
      \State \quad \textbf{case} $s :: \pi'$ $\rightarrow$ $(\neg s :: \pi') :: \textsc{Negate}(\pi')$
    \EndFunction
  \end{algorithmic}
\end{algorithm}

\renewcommand{\magic}{\hspace{-16em}}

\begin{figure}[p]\footnotesize

\judgement{Staged Concolic Continuation Semantics}{}

  \begin{alignat*}{3}
    & \SConcdeno{\cdot} : \mathrlap{\Typ{List[Inst]} \rightarrow \HL{\QuoteT{\Typ{Memo}}}} \\
    & \hspace{3.5em} \mathrlap{\rightarrow (\HL{\QuoteT{\Typ{Stack^\#}}} \times \HL{\QuoteT{\Typ{Env^\#}}}  \times \HL{\QuoteT{\Typ{PC^\#}}} \times \HL{\QuoteT{\Typ{Cont^\#}}} \times \HL{\Typ{Trail^\#}}) \rightarrow \HL{\QuoteT{\Typ{Ans}}}} \\
    & \SConcdeno{\mathit{nil}}(\DV{\memo})(\DV{\sigma^\#}, \DV{\rho^\#}, \DV{\pi^\#}, \DV{\kappa^\#}, \SV{\theta^\#}) && = \Quote{\Splice{\DV{\kappa^\#}}(\Splice{\DV{\sigma^\#}}, \Splice{\DV{\rho^\#}}, \Splice{\DV{\pi^\#}})} \\
    & \SConcdeno{\Inst{nop} :: \SV{\mathit{rest}}}(\DV{\memo})(\DV{\sigma^\#}, \DV{\rho^\#}, \DV{\pi^\#}, \DV{\kappa^\#}, \SV{\theta^\#}) && =
      \SConcdeno{\SV{\mathit{rest}}}(\DV{\memo})({\DV{\sigma^\#}}, \DV{\rho^\#}, \DV{\pi^\#}, \DV{\kappa^\#}, \SV{\theta^\#}) \\
    & \SConcdeno{t.\Inst{const} ~ \SV{c} :: \SV{\mathit{rest}}}(\DV{\memo})(\DV{\sigma^\#}, \DV{\rho^\#}, \DV{\pi^\#}, \DV{\kappa^\#}, \SV{\theta^\#}) && =\\
       & && \magic
      \SConcdeno{\SV{\mathit{rest}}}(\DV{\memo})(\Quote{\la\Splice{\lift(\SV{c})}, \Splice{\lift(\SV{c})}\ra :: \Splice{\DV{\sigma^\#}}}, \DV{\rho^\#}, \DV{\pi^\#}, \DV{\kappa^\#}, \SV{\theta^\#}) \\
    & \SConcdeno{t.\Inst{add} :: \SV{\mathit{rest}}}(\DV{\memo})(\DV{\sigma^\#}, \DV{\rho^\#}, \DV{\pi^\#}, \DV{\kappa^\#}, \SV{\theta^\#}) && =\\
      & && \magic \Quote{\Typ{let} ~ \la v_1, s_1\ra \coloneqq \Splice{\DV{\sigma^\#}.\mathsf{head}} ~ \Typ{in} \\
      & && \magic \qspace\Typ{let} ~ \la v_2, s_2\ra \coloneqq \Splice{\DV{\sigma^\#}.\mathsf{tail}.\mathsf{head}} ~ \Typ{in} \\
      & && \magic \qspace\Splicee{\SConcdeno{\SV{\mathit{rest}}}(\DV{\memo})(\Quote{\la v_1 + v_2, s_1 ~ +^\# ~ s_2\ra :: \Splice{\DV{\sigma^\#}}.\mathsf{tail}.\mathsf{tail}}, \DV{\rho^\#}, \DV{\pi^\#}, \DV{\kappa^\#}, \SV{\theta^\#})}} \\
    & \SConcdeno {\Inst{local.get} ~ \SV{x} :: \SV{\mathit{rest}}} (\DV{\memo}) (\DV{\sigma^\#}, \DV{\rho^\#}, \DV{\pi^\#}, \DV{\kappa^\#}, \SV{\theta^\#}) && =
      \SConcdeno{\SV{\mathit{rest}}}(\DV{\memo})(\Quote{\Splice{\DV{\rho^\#}(\Splicee{\lift(\SV{x})})} :: \Splice{\DV{\sigma^\#}}}, \DV{\rho^\#}, \DV{\pi^\#}, \DV{\kappa^\#}, \SV{\theta^\#}) \\
    & \SConcdeno{\Inst{local.set} ~ \SV{x} :: \SV{\mathit{rest}}}(\DV{\memo})(\DV{\sigma^\#}, \DV{\pi^\#}, \DV{\rho^\#}, \DV{\kappa^\#}, \SV{\theta^\#}) && = \\
      & && \magic \Quote{\Typ{let} ~ \la v, s\ra \coloneqq \Splice{\DV{\sigma^\#}.\mathsf{head}} ~ \Typ{in} \\
      & && \magic \qspace\Splicee{\SConcdeno{\SV{\mathit{rest}}}(\DV{\memo})(\Quote{\Splice{\DV{\sigma^\#}}.\mathsf{tail}}, \Quote{\Splice{\DV{\rho^\#}}[\Splicee{\lift(\SV{x})} \mapsto \la v, s\ra]}, \DV{\pi^\#}, \DV{\kappa^\#}, \SV{\theta^\#})}} \\
    & \SConcdeno{\Inst{block} ~ (t^m \to t^n) ~ \SV{\es} :: \SV{\mathit{rest}}}(\DV{\memo})(\DV{\sigma^\#}, \DV{\rho^\#}, \DV{\pi^\#}, \DV{\kappa^\#}, \SV{\theta^\#}) && = \\
      & && \magic \Quote{\Typ{let} ~ \kappa^\#_1 \coloneqq \lambda (\sigma^\#_1, \rho^\#_1, \pi^\#_1). \Splicee{\SConcdeno{\SV{\mathit{rest}}}(\Quote{\truncate{\sigma^\#_1}{n} \mathbin{+\!\!+} \Splice{\DV{\sigma^\#}}.\pop_m}, \Quo{\rho^\#_1}, \Quo{\pi^\#_1}, \DV{\kappa^\#}, \SV{\theta^\#})} ~ \Typ{in} \\
      & && \magic \qspace\Splicee{\SConcdeno{\SV{\es}}(\Quote{\Splice{\DV{\sigma^\#}}.\take_m}, \DV{\rho^\#}, \DV{\pi^\#}, \Quo{\kappa^\#_1}, \Quo{\kappa^\#_1} :: \SV{\theta^\#})} } \\
    & \SConcdeno{\Inst{loop} ~ (t^m \to t^n) ~ \SV{\es} :: \SV{\mathit{rest}}}(\DV{\memo})(\DV{\sigma^\#}, \DV{\rho^\#}, \DV{\pi^\#}, \DV{\kappa^\#}, \SV{\theta^\#}) && = \\
      & && \magic \Quote{\Typ{let} ~ \kappa^\#_1 \coloneqq \lambda (\sigma^\#_1, \rho^\#_1, \pi^\#_1). \Splicee{\SConcdeno{\SV{\mathit{rest}}}(\DV{\memo})(\Quote{\truncate{\sigma^\#_1}{n} \mathbin{+\!\!+} \Splice{\DV{\sigma^\#}}.\pop_m}, \Quo{\rho^\#_1}, \Quo{\pi^\#_1}, \DV{\kappa^\#}, \SV{\theta^\#})} ~ \Typ{in} \\
       & && \magic \qspace \Typ{fix} ~ \kappa^\#_2 \coloneqq \lambda (\sigma^\#_2, \rho^\#_2, \pi^\#_2). \Splicee{\SConcdeno{\SV{\es}}(\DV{\memo})(\Quote{\truncate{\sigma^\#_2}{m}}, \Quo{\rho^\#_2}, \Quo{\pi^\#_2}, \Quo{\kappa^\#_1}, \Quo{\kappa^\#_2} :: \SV{\theta^\#})} ~ \Typ{in} \\
       & && \magic \qspace \kappa^\#_2(\Splice{\DV{\sigma^\#}}.\take_m , \Splice{\DV{\rho^\#}}, \Splice{\DV{\pi^\#}}) } \\
    & \SConcdeno{\Inst{if} ~ (t^m \to t^n) ~ \SV{\es_1} ~ \SV{\es_2} :: \SV{\mathit{rest}}}(\DV{\memo})(\DV{\sigma^\#}, \DV{\rho^\#}, \DV{\pi^\#}, \DV{\kappa^\#}, \SV{\theta^\#}) && = \\
       & && \magic \Quote{\Typ{let} ~ \la v, s\ra \coloneqq \Splice{\DV{\sigma^\#}.\mathsf{head}} ~ \Typ{in} \\
       & && \magic \qspace\Typ{let} ~ \sigma^\#_{\mathit{arg}} \coloneqq \Splice{\DV{\sigma^\#}}.\mathsf{tail}.\take_m ~ \Typ{in} \\
       & && \magic \qspace\Typ{let} ~ \kappa^\#_1 \coloneqq \lambda (\sigma^\#_1, \rho^\#_1, \pi^\#_1). \Splicee{\SConcdeno{\SV{\mathit{rest}}}(\DV{\memo})(\Quote{\truncate{\sigma^\#_1}{n} \mathbin{+\!\!+} \Splice{\DV{\sigma^\#}}.\tail.\pop_m}, \Quo{\rho^\#_1}, \Quo{\pi^\#_1}, \DV{\kappa^\#}, \SV{\theta^\#})} ~ \Typ{in} \\
       & && \magic \qspace \Typ{let} ~ \pi^\#_1 \coloneqq \Typ{if} ~ v \equiv 0 ~ \Typ{then} ~ \neg s :: \Splice{\DV{\pi^\#}} ~ \Typ{else} ~ s :: \Splice{\DV{\pi^\#}} ~ \Typ{in} \\
       & && \magic \qspace \Typ{if} ~ v \equiv 0 ~ \Typ{then} ~ \Splicee{\SConcdeno{\SV{\es_1}}(\DV{\memo})(\Quo{\sigma^\#_{\mathit{arg}}}, \DV{\rho^\#}, \Quo{\pi^\#_1}, \Quo{\kappa^\#_1}, \Quo{\kappa^\#_1} :: \SV{\theta^\#})} ~ \\
       & && \hspace{-12.75em} \qspace \Typ{else} ~ \Splicee{\SConcdeno{\SV{\es_2}}(\DV{\memo})(\Quo{\sigma^\#_{\mathit{arg}}}, \DV{\rho^\#}, \Quo{\pi^\#_1}, \Quo{\kappa^\#_1}, \Quo{\kappa^\#_1} :: \SV{\theta^\#})}
        } \\
    & \SConcdeno{\Inst{br} ~ \SV{\ell} :: \SV{\mathit{rest}}}(\DV{\memo})(\DV{\sigma^\#}, \DV{\rho^\#}, \DV{\pi^\#}, \DV{\kappa^\#}, \DV{\theta^\#}) && = \Quote{\Splicee{\SV{\theta^\#}(\SV{\ell})}(\Splice{\DV{\sigma^\#}}, \Splice{\DV{\rho^\#}}, \Splice{\DV{\pi^\#}})} \\
    & \SConcdeno{\Inst{call} ~ \SV{x} :: \SV{\mathit{rest}}}(\DV{\memo})(\DV{\sigma^\#}, \DV{\rho^\#}, \DV{\pi^\#}, \DV{\kappa^\#}, \DV{\theta^\#}) && = \\
       & && \magic \Typ{let} ~ \{ \Typ{type}: t^m \to t^n, \Typ{locals}: \SV{\mathit{ts}}, \Typ{body}: \SV{\es} \} \coloneqq \Typ{lookupFunc}(\SV{x}) ~ \Typ{in} \\
       & && \magic \Quote{\Typ{let} ~ f \coloneqq \Splice{\DV{\memo}}(\Splicee{\lift(\SV{x})}) ~ \Typ{in} \\
       & && \magic \qspace \Typ{let} ~ \rho^\#_1 \coloneqq \mathsf{buildEnv}(\Splice{\DV{\sigma^\#}.\take_m}, \Splicee{\lift(\SV{\mathit{ts}})}) \\
       & && \magic \qspace\Typ{let} ~ \kappa^\#_1 \coloneqq \lambda (\sigma^\#_1, \rho^\#_1, \pi^\#_1). \Splicee{\SConcdeno{\SV{\mathit{rest}}}(\DV{\memo})(\Quote{\truncate{\sigma^\#_1}{n} \mathbin{+\!\!+} \Splice{\DV{\sigma^\#}}.\pop_m}, \DV{\rho^\#_1}, \DV{\pi^\#_1}, \DV{\kappa^\#}, \SV{\theta^\#})} ~ \Typ{in} \\
       & && \magic \qspace\Splice{\DV{f}}(\rho^\#_1, \Splice{\DV{\pi^\#}}, \kappa^\#_1, \Splice{\DV{\memo}}) } \\
    & \SConcdeno{\Inst{return} :: \SV{\mathit{rest}}}(\DV{\memo})(\DV{\sigma^\#}, \DV{\rho^\#}, \DV{\pi^\#}, \DV{\kappa^\#}, \SV{\theta^\#}) && = \Quote{\Splicee{\SV{\theta^\#}.\mathsf{last}}(\Splice{\DV{\sigma^\#}}, \Splice{\DV{\rho^\#}}, \Splice{\DV{\pi^\#}})}
  \end{alignat*}
  \caption{The staged concolic semantics of \lang (complete version of
    \Cref{fig:staging:semantics}).}\label{fig:staging:semantics:full} 
\end{figure}

\begin{algorithm}[t]
  \small
  \caption{Concolic Scheduler Working with Staged Concolic Semantics}
  \label{fig:staging:driver}
  \begin{algorithmic}[1]
    \Function{\textsc{Scheduler}}{$mod \in \Typ{Module}$, $f \in \Typ{Identifier}$, $\mathit{input} \in \Typ{List[Symbolic]}$}
      \State $WL \gets \{\, [] \, \}$ \Comment{$WL \in \Pow{\Typ{PC^\#}}$}
      \State $Seen \gets \varnothing$ \Comment{$Seen \in \Pow{\Typ{PC^\#}}$}
      \State $\HL{\mathit{memo} \gets \mathsf{run}~\MConcdeno{mod}}$
      \State $\HL{\mathit{entry} \gets \mathit{memo}(f)}$
      \While{$WL \neq \varnothing$}
        \State pop $\pi$ from $WL$
        \IIf{$\pi \in Seen$} \textbf{continue}
        \State $model \gets \mathit{solve}(\pi)$
        \IIf{$model = \UNSAT$} \textbf{continue}
        \State $\pi_1 \gets \HL{\mathit{entry}(\mathit{instantiate}(\mathit{input}, model), [], \kappa^\#_{halt}, \mathit{memo})}$
        \State $Seen \gets Seen \cup \{\pi_1\}$
        \ForAll{$\pi' \in \textsc{Negate}(\pi_1)$}
          \State $WL \gets WL \cup \{\pi'\}$
        \EndFor
      \EndWhile
    \EndFunction
  \end{algorithmic}
\end{algorithm}

\begin{algorithm}[t]
  \small
  \caption{Concolic Snapshot Scheduler (complete version of
    \Cref{fig:snapshot:driver})}
  \label{fig:snapshot:driver:full}
  \begin{algorithmic}[1]
    \Function{\textsc{Scheduler}}{$mod \in \Typ{Module}$, $f \in \Typ{Identifier}$, $\mathit{input} \in \Typ{List[Symbolic]}$}
    \State $\mathit{memo} \gets \mathsf{run} ~ \MConcdeno{mod}$
    \State $\mathit{entry} \gets \mathit{memo}(f)$
    \State $WL \gets \HL{\{\, \la [], \la \lambda () . \mathit{entry}(\kappa^\#_{halt}, \mathit{memo}), \mathit{input}, []\ra\ra \, \}}$ \Comment{$\HL{WL \in \Pow{\Typ{List[Symbolic]} \times \Typ{Snapshot}}}$}
    \State $Seen \gets \varnothing$ \Comment{$\HL{Seen \in \Pow{\Typ{List[Symbolic]}}}$}
      \While{$WL \neq \varnothing$}
        \State pop $\la\pi^\#, \HL{\la\kappa, \sigma^\#, \rho^\#\ra}\ra$ from $WL$
        \IIf{$\pi^\# \in Seen$} \textbf{continue}
        \State $m \gets \mathit{solve}(\pi^\#)$
        \IIf{$m = \UNSAT$} \textbf{continue}
        \State $\sigma_g^\# \texttt{:=} \mathit{instantiate}(\sigma^\#, m)$ \Comment{update global symbolic stack}
        \State $\rho_g^\# \texttt{:=} \mathit{instantiate}(\rho^\#, m)$ \Comment{update global symbolic environment}
        \State $\pi_g^\# \texttt{:=} \pi^\#$ \Comment{update global path condition}
        \State $\HL{\Pi^\# \gets \kappa()}$
        \State $Seen \gets Seen \cup \{\HL{\Pi^\#.\mathsf{pc}}\}$ \Comment{$\Pi^\#.\mathsf{pc}$ extracts a list of symbolic conditions}
        \ForAll{$\HL{\la\pi_1, \Sigma_1\ra}\in \textsc{Negate}(\Pi^\#)$}
          \State $WL \gets WL \cup \{\HL{\la\pi_1, \Sigma_1\ra}\}$
        \EndFor
      \EndWhile
    \EndFunction
    \\
    \Function{\textsc{Negate}}{$\Pi^\# \in \Typ{PC^\#}$} $\rightarrow \HL{\Typ{List}[\Typ{List[Symbolic]} \times \Typ{Snapshot}]}$
      \State \textbf{match} $\Pi^\#$ \textbf{with}
      \State \quad \textbf{case} $[]$ $\rightarrow$ $[]$
      \State \quad \textbf{case} $\la s, \Sigma\ra :: \Pi'$ $\rightarrow$ $\la\neg s :: \Pi'.\mathsf{pc}, \Sigma\ra :: \textsc{Negate}(\Pi')$
    \EndFunction
  \end{algorithmic}
\end{algorithm}

\begin{algorithm}[t]
  \small
  \caption{Concolic Snapshot Scheduler with Heuristic}
  \label{fig:heuristic:driver}
  \begin{algorithmic}[1]
    \Function{\textsc{Scheduler}}{$mod \in \Typ{Module}$, $f \in \Typ{Identifier}$, $\mathit{input} \in \Typ{List[Symbolic]}$}
      \State $WL \gets \{\, [] \, \}$
      \State $Seen \gets \varnothing$
      \State $\mathit{defs} \gets \ODeno{\MConcdeno{mod}}$
      \State $\mathit{entry} \gets \mathit{defs}(f)$
      \State $\kappa^\#_{halt} \gets \ODeno{\lift(\kappa^\#_{halt})}$
      \While{$WL \neq \varnothing$}
        \State pop $\la\pi^\#, \la\kappa, \rho^\#, \sigma^\#, \HL{c}\ra\ra$ from $WL$
        \IIf{$\pi^\# \in Seen$} \textbf{continue} 
        \State $m \gets \mathit{solve}(\pi)$
        \IIf{$m = \UNSAT$}
          \textbf{continue}
        \If{$\HL{c / (\delta * (\mathsf{sizeOfSymExp}(\sigma^\#) +
          \mathsf{sizeOfSymExp}(\rho^\#))) > 1}$}
          \State $\HL{\sigma_g^\# \texttt{:=} \mathit{instantiate}(\sigma^\#, m)}$
          \State $\HL{\rho_g^\# \texttt{:=} \mathit{instantiate}(\rho^\#, m)}$
          \State $\HL{\pi_g^\# \texttt{:=} \pi^\#}$
          \State $\HL{\Pi^\# \gets \kappa()}$
        \Else
          \State $\HL{\Pi^\# \gets \mathit{entry}(\mathit{instantiate}(\mathit{input}), [], [], \kappa^\#_{halt})}$
        \EndIf
        \State $Seen \gets Seen \cup \{\Pi^\#.\mathit{pc}\}$
        \ForAll{$\la\pi_1, \Sigma_1\ra\in \textsc{Negate}(\Pi^\#)$}
          \State $WL \gets WL \cup \{\la\pi_1, \Sigma_1\ra\}$
        \EndFor
      \EndWhile
    \EndFunction
  \end{algorithmic}
\end{algorithm}

\clearpage

\end{document}